\documentclass{aa}
\usepackage{epsfig}

\catcode`\@=11
\def\gsim{\ifmmode{\mathrel{\mathpalette\@versim>}}
    \else{$\mathrel{\mathpalette\@versim>}$}\fi}
\def\lsim{\ifmmode{\mathrel{\mathpalette\@versim<}}
    \else{$\mathrel{\mathpalette\@versim<}$}\fi}
\def\@versim#1#2{\lower 2.9truept \vbox{\baselineskip 0pt \lineskip
    0.5truept \ialign{$\m@th#1\hfil##\hfil$\crcr#2\crcr\sim\crcr}}}
\catcode`\@=12
\def\Rt{R_{\rm t}}
\def\Re{R_{\rm e}}
\def\Reo{R_{\rm e}^0}
\def\ellef{\ell_{\rm e}}
\def\Ie{I_{\rm e}}

\def\Sigt{\Sigma_{\theta}}
\def\Sigtl{\Sigt(\ell)}
\def\Sigtlp{\Sigt(\ell\prime)}
\def\Mt{M_{\theta}}
\def\Mtl{\Mt (\ell)}
\def\qt{q(\theta)}
\def\betaf{{\rm B}}
\def\hypgeof{_2{\rm F}_1}
\def\Lz{L_z}

\def\los{{\it los }}
\def\xv{{\bf x}}
\def\dtx{d^3\xv}
\def\xci{x_i}
\def\xcj{x_j}
\def\zt{z_{\rm t}}
\def\xp{x\prime}
\def\yp{y\prime}
\def\zp{z\prime}
\def\zpp{\zp_+}
\def\zpm{\zp_-}

\def\nv{{\bf n}}
\def\nci{n_i}
\def\ncj{n_j}
\def\ncu{n_1}
\def\nct{n_3}
\def\vv{{\bf v}}
\def\dtvv{d^3\vv}
\def\ttv{{\rm v}}
\def\tvx{\ttv_x}
\def\tvy{\ttv_y}
\def\tvz{\ttv_z}
\def\tvphi{\ttv_{\varphi}}
\def\tvphim{\overline{\ttv}_{\varphi}}
\def\tvphis{\tvphi^2}
\def\tvphism{\overline{\tvphis}}
\def\tvR{\ttv_R}
\def\vcx{v_x}
\def\vcy{v_y}
\def\vcz{v_z}
\def\vci{v_i}
\def\vcj{v_j}
\def\vcR{v_R}
\def\vcphi{v_{\varphi}}
\def\vn{v_{\rm n}}
\def\vp{v_{\rm p}}
\def\Vp{V_{\rm p}}
\def\Vps{\Vp^2}
\def\sigc{\sigma_0}
\def\sigR{\sigma_R}
\def\sigRs{\sigR^2}
\def\sigz{\sigma_z}
\def\sigzs{\sigz^2}
\def\sigxx{\sigma_{xx}}
\def\sigyy{\sigma_{yy}}
\def\sigzz{\sigma_{zz}}
\def\sigij{\sigma_{ij}}
\def\sigphi{\sigma_{\varphi}}
\def\sigphis{\sigphi^2}
\def\sign{\sigma_{\rm n}}
\def\signs{\sign^2}
\def\sigp{\sigma_{\rm p}}
\def\sigps{\sigp^2}
\def\sigob{\sigma_{\rm los}}
\def\sigobm{\sigma_{\rm los,max}}
\def\sigobs{\sigob^2}
\def\sigobsal{\sigma_{\rm los,a}^2(\ell)}
\def\sigoba{\sigma_{\rm los,a}}
\def\Rtild{\widetilde R}
\def\ztild{\widetilde z}

\def\ltild{\widetilde\ell}
\def\xpt{\widetilde\xp}
\def\ypt{\widetilde\yp}
\def\zpt{\widetilde\zp}

\def\rhoc{\rho_0}
\def\rhonijq{\rho_{nij}(q)}
\def\phinijq{\phi_{nij}(q)}
\def\zetanijklq{\zeta_{nijkl}(q)}
\def\etanijklq{\eta_{nijkl}(q)}
\def\Wij{W_{ij}}
\def\Wuu{W_{11}}
\def\Wdd{W_{22}}
\def\Wtt{W_{33}}
\def\wz{w_0}
\def\wu{w_1}
\def\wt{w_3}
\def\ser{R^{1/m}}

\def\dfp{\Delta{\rm FP}}
\def\Refp{R_{\rm e,FP}}
\def\delinp{\delta_{\rm inp}}
\def\onesigproj{\sigma_{\rm proj}}
\def\onesigphys{\sigma_{\rm phys}}
\def\onesigint{\sigma_{\rm int}}
\def\onesiginp{\sigma_{\rm inp}}
\def\rms{{\it rms}}
\def\ml{\Upsilon_*}
\def\Kvir{K_{\rm V}}

\begin{document}
   \title{Projection effects on the FP thickness:}

   \subtitle{a Monte-Carlo exploration}

   \author{B. Lanzoni\inst{1}  \and 
           L. Ciotti\inst{2,3,4}}
           
   \offprints{lanzoni@bo.astro.it}

   \institute{INAF - Osservatorio Astronomico di Bologna, via Ranzani 1,
              I-40127 Bologna, Italy
         \and
              Dipartimento di Astronomia, 
              Universit\`a di Bologna, via Ranzani 1,
              I-40127 Bologna, Italy
         \and
              Scuola Normale Superiore, Piazza dei Cavalieri 7,
              I-56126 Pisa, Italy
         \and
              Princeton University Observatory, Peyton Hall, Princeton
              NJ 08544-1001, USA} 

   \date{Accepted: March 24, 2003. Submitted in original form: October
   29, 2002}

\abstract{We study the contribution of projection effects to the
intrinsic thickness of the Fundamental Plane (FP) of elliptical
galaxies. The Monte--Carlo mapping technique between model properties
and observed quantities, introduced by Bertin, Ciotti, \& Del Principe
(2002), is extended to oblate, two--integrals galaxy models, with
non--homologous density profiles, adjustable flattening, variable
amount of ordered rotational support, and for which all the relevant
projected dynamical quantities can be expressed in fully analytical
way. In agreement with previous works, it is found that projection
effects move models not exactly parallel to the edge--on FP, by an
amount that can be as large as the observed FP thickness. The
statistical contribution of projection effects to the FP thickness is
however marginal, and the estimated physical FP $rms$ thickness is
$\simeq 90$\% of the observed one (when corrected for measurement
errors).   }

\authorrunning{Lanzoni and Ciotti}
\titlerunning{Projection effects on the FP thickness}
\maketitle

\section{Introduction}

In the observational three--dimensional space of central velocity
dispersion $\sigc$, (circularized) effective radius $\Re$, and mean
surface brightness within the effective radius $\Ie$, early--type
galaxies approximately locate on a plane, called the Fundamental Plane 
(hereafter FP; Dressler et al. 1987; Djorgovski \& Davis 1987), and
represented by the best--fit relation:
\begin{equation}
\log\Re =\alpha\log\sigc +\beta\log\Ie +\gamma.
\label{eq:FP}
\end{equation}
The coefficients $\alpha$, $\beta$, and $\gamma$ depend slightly on
the considered photometric band (e.g., Pahre, de Carvalho, \&
Djorgovski, 1998; Scodeggio et al. 1998). By measuring $\Re$ in kpc,
$\sigc$ in km/s, and $\Ie = L/(2\pi\Re^2)$ in $L_\odot/$pc$^2$ (where
$L$ is the total galaxy luminosity), reported values in the Gunn r
band are $\alpha =1.24\pm 0.07$, $\beta =-0.82\pm 0.02$, $\gamma
=0.182$\footnote{This value of $\gamma$ refers to the Coma cluster and
to $H_0=50$ km s$^{-1}$ Mpc$^{-1}$.} (J{\o}rgensen, Franx, \&
Kj{\ae}rgaard 1996, hereafter JFK96).  One of the most striking
observational properties of the FP is its small and nearly constant
scatter: the distribution of $\log\Re$ around the best--fit FP has a
measured $\rms$ (after correction for measurement errors, hereafter
$\onesigint$), that corresponds to a scatter in $\Re$ at fixed $\sigc$
and $\Ie$ ranging from 15\% to 20\% (see, e.g., Faber et al. 1987;
JFK96).
% In this paper we assume as reference
% value $\onesigint = 0.057$, the value quoted by JFK96, for the galaxies
% with velocity dispersion larger than 100 km/s.

For a stationary stellar system the scalar virial theorem can be
written as
\begin{equation}
{G\ml L\over\Re}=\Kvir\sigc^2,
\label{eq:VT}
\end{equation}
where $\ml$ is the {\it stellar} mass--to--light ratio in the
photometric band used for the determination of $L$ and $\Re$, while
the coefficient $\Kvir$ takes into account projection effects, 
the specific mass density, the stellar orbital distribution (such as 
velocity dispersion anisotropy and rotational support), and the
effects related to the presence of dark matter.  Equations (\ref{eq:FP}) and
(\ref{eq:VT}) imply that in real galaxies, no matter how complex their
structure is, $\ml/\Kvir$ is a well--defined function of any two of the
three observables $(L,\Re,\sigc)$.  For example, by eliminating
$\sigc$ from eqs. (\ref{eq:FP}) and (\ref{eq:VT}) one obtains that 
along the FP
\begin{equation}
{\ml\over\Kvir}\propto \Re^{(2-10\beta+\alpha)/\alpha}
                        L^{(5\beta-\alpha)/\alpha}, 
\label{eq:mlkvir}
\end{equation}
where the dependence of the ratio $\ml/\Kvir$ on galaxy properties is
commonly referred as the ``FP tilt''. The physical content of
eq. (\ref{eq:mlkvir}) is truly remarkable: all stellar systems
described by eq. (\ref{eq:VT}) are in virial equilibrium, but only
those for which $\ml/\Kvir$ scales according to eq. (\ref{eq:mlkvir})
(and with the same scatter) correspond to real galaxies.  In other
words, eq. (\ref{eq:mlkvir}) indicates that {\it structural/dynamical}
($\Kvir$) and {\it stellar population} ($\ml$) properties in real
galaxies are strictly connected, possibly as a consequence of their
formation process: understanding the origin of the FP tilt is thus of
the utmost importance for the understanding of galaxy formation.

A first possibility in this direction is to focus on the
variation of a {\it single} galaxy property among the plethora in
principle appearing in the quantity $\ml/\Kvir$, while fixing all the
others to some prescribed value: we call this approach {\it orthogonal
exploration} of the parameter space.  For instance, one can explore
the possibility that a systematic variation of $\ml$ with $L$ is at
the origin of the FP tilt, while considering the galaxies as strictly
homologous systems (i.e., with density and dynamical structures only
differing for the physical scales, and thus $\Kvir=$const. See, e.g.,
Bender, Burstein, \& Faber 1992; Renzini \& Ciotti 1993; van Albada,
Bertin \& Stiavelli 1995, hereafter vABS; Prugniel \& Simien 1996).
Another possibility is to enforce a constant $\ml$, and to assume that
the galaxy density profiles, dark matter content and distribution,
stellar orbital distribution, and so on, vary systematically with $L$
(see, e.g., Ciotti \& Pellegrini 1992; Caon, Capaccioli \& D'Onofrio
1993; Renzini \& Ciotti 1993; Djorgovski 1995; Hjorth \& Madsen 1995;
Ciotti, Lanzoni \& Renzini 1996, hereafter CLR; Graham \& Colless
1997; Ciotti \& Lanzoni 1997; Prugniel \& Simien 1997).

Orthogonal explorations lead to important results, but, besides
starting from a (more or less) well motivated choice of the specific
parameter assumed to be responsible for the FP tilt, they also bring
to a {\it fine tuning} problem: the large variation of such a
parameter along the FP, necessary to reproduce the tilt, must be
characterized by a small scatter of it at any fixed position on
the FP in order to preserve the observed small thickness (e.g.,
Renzini \& Ciotti 1993; CLR).  Moreover, the severity of the fine
tuning problem is strengthened by the unavoidable projection effects
associated with the three--dimensional shape of galaxies, if they also
contribute to FP thickness.  Thus, {\it the interpretation of the FP
cannot be limited to the study of its tilt only, but requires to take
consistently into account also its thinness}.

Recently, a statistical approach to this problem, based on
Monte--Carlo simulations and overcoming the intrinsic limitations of
orthogonal explorations has been proposed (Bertin, Ciotti \& Del
Principe 2002, hereafter BCD). In this study the authors showed that,
ascribing the origin of the FP tilt to the {\it combined} effect of
luminosity dependent mass--to--light ratio and shape parameter $m$ in
spherically symmetric and isotropic $\ser$ models (Sersic 1968), can
reconcile the FP tilt with the observed large dispersion of $m$ at
fixed galaxy luminosity (see Figs. 5 and 6 in CLR and Figs. 7 and 9 in
BCD).  Note, however, that in the BCD analysis the FP thickness is
entirely produced by variations from galaxy to galaxy of their {\it
physical} properties , as a consequence of the assumption of spherical
symmetry.  On the other hand, elliptical galaxies are in general non
spherical, and the quantities entering the FP expression do depend on
the observation angle: it is therefore of great interest to estimate
the contribution of projection effects to the FP thickness, and to
quantify its {\it physical} scatter.  Few analytical works have
addressed this issue in the past (e.g., Faber et al. 1987; Saglia,
Bender \& Dressler 1993; J{\o}rgensen, Franx, \& Kj{\ae}rgaard 1993;
Prugniel \& Simien 1994; JFK96; vABS), their conclusions pointing in
the direction of a small contribution of projection effects to the FP
thickness.  A different source of information on projection effects is
also represented by the end--products of N--body numerical simulations
(see, e.g., Pentericci, Ciotti, \& Renzini 1995; Nipoti, Londrillo \&
Ciotti 2002ab, 2003; Gonz\'ales \& van Albada 2002).  The impression
one gets from these simulations is that projection effects can be
significant contributors to the FP thickness, the range spanned by the
models for changing viewing angle being comparable to $\onesigint$ or
more.

We explore this matter further, by extending the BCD approach to a
class of oblate ellipsoids with non homologous density profiles,
adjustable flattening and variable amount of internal velocity
streaming.  However, in order to maintain the dimension of the
parameter space acceptable we do not take into account the presence of
DM halos, and the stellar mass--to--light ratio $\ml$ is assumed to be
constant within each galaxy.  For these models all the relevant
quantities can be expressed explicitly, thus allowing for fast
numerical calculation.  The paper is organized as follows.  In Section
2 we derive the relevant properties of the adopted models.  In Section
3 we illustrate in detail a few representative cases, focusing on the
effects of the various model parameters on the observational
properties entering the FP relation.  In Section 4 the results of the
Monte--Carlo investigations are shown, and finally, in Section 5 we
summarize and discuss the results.  Appendix A collects the explicit
formulas describing the model internal dynamics, while in Appendix B
we derive the expressions for the associated projected quantities. In
Appendix C the simplest model of the family (the homogeneous
ellipsoid) is described in detail.

\section{The models}

\subsection{3D quantities}

In our study we use a family of oblate galaxy models with homeoidal
density distribution, belonging to the so--called {\it Ferrers
ellipsoids} (Ferrers 1877). The density profile is given by
\begin{equation}
\rho(m) = \rhoc\times \cases{(1-m^2)^n &if $0\leq m\leq 1$,\cr 
                                            0 &if $m>1$,\cr}
\label{eq:rho}
\end{equation}
where $\rhoc$ is the central density, $n\geq 0$ is an integer number,
and in cylindrical coordinates\footnote{These coordinates are related
to the natural Cartesian coordinate system by the relations
$R=\sqrt{x^2 + y^2}$, $\cos\varphi=x/R$, $\sin\varphi=y/R$.}
$(R,\varphi ,z)$ the isodensity surfaces are labeled by $m^2 \equiv
R^2/\Rt^2 + z^2/(q^2\Rt^2)$.  With this choice $\Rt$ is the model
semi--major axis, while its flattening is given by $0<q\leq 1$. Note
that these density profiles, when considered in detail, are only a
rough approximation of those of real galaxies, especially for low
values of $n$.  {\it However, most of the model properties that are
relevant for this study show a behavior surprisingly similar to that
of galaxy models with more realistic density profiles} (see Sections
3).  In addition the above mentioned properties can be explicitly
written in analytic form, making the models suitable for Monte--Carlo
simulations.

The mass within $m$ and the total mass of the models are given by
\begin{equation}
M(m) = \rhoc\Rt^3 2\pi q\betaf_{3/2,n+1}(m^2),
\label{eq:mass}
\end{equation}
and
\begin{equation}
M = \rhoc\Rt^3 2\pi q\betaf\left({3\over 2},n+1\right),
\label{eq:masstot}
\end{equation}
respectively, where $\betaf_{a,b}(z)\equiv \int_0^z t^{a-1}\,
(1-t)^{b-1}dt$ is the incomplete Euler Beta function,
$\betaf(a,b)\equiv\betaf_{a,b}(1)=\Gamma(a)\Gamma(b)/\Gamma(a+b)$ is
the complete Euler Beta function, and $\Gamma$ is the complete Euler
Gamma function.

We assume that the density profiles in eq. (\ref{eq:rho}) are
supported by a dynamics described by a two--integrals distribution
function $f=f(E,\Lz)$ (where $E$ and $\Lz$ are the energy and the $z$
component of the angular momentum of stars). Thus, the Jeans equations
reduce to\footnote{We use symbol $\vv$ for the velocity in the phase
space, while $\vec v(\xv)\equiv \overline{\vv}$ is the {\it streaming}
velocity as defined in eq. (B2).  In general, a bar over a quantity
means average over phase--space velocities.}
\begin{equation}
{\partial\rho\sigRs\over\partial z} = -\rho {\partial\phi\over\partial z}, 
\label{eq:jeans1}
\end{equation}
and
\begin{equation}
{\partial\rho\sigRs\over\partial R}-{\rho (\tvphism -\sigRs)\over R} = 
-\rho {\partial\phi\over\partial R},
\label{eq:jeans2}
\end{equation}
where $\phi$ is the gravitational potential, $\vcR = \vcz = 0$
everywhere, the off--diagonal elements of the velocity dispersion
tensor vanish, and $\sigRs = \sigzs$ (see, e.g., Binney \& Tremaine
1987, hereafter BT).  The appropriate boundary conditions are $\sigRs
= \sigzs = 0$ on $m=1$ (Ciotti 2000), and so, the formal solution of
eqs. (\ref{eq:jeans1})-(\ref{eq:jeans2}) is:
\begin{equation}
\rho\sigRs =\int_z^{\zt}\rho {\partial\phi\over\partial z}dz , 
\label{eq:soljeans1}
\end{equation}
where $\zt\equiv q\sqrt{\Rt^2 -R^2}$, and
\begin{equation}
\rho(\tvphism -\sigRs) = R\left(
{\partial\rho\sigRs\over \partial R} + \rho
{\partial\phi\over\partial R}\right).
\label{eq:soljeans2}
\end{equation}
As it is well known, the gravitational potential of homeoidal systems
can be obtained by evaluating a two--dimensional integral (see, e.g.,
Chandrasekhar 1969), but in general this integral cannot be expressed
in terms of elementary functions.  From this point of view the density
profiles adopted here are a nice exception: their potential can be
written explicitly (for $n$ integer) as a finite sum of integer powers
of $R$ and $z$. Thus, from eqs. (\ref{eq:soljeans1}) and
(\ref{eq:soljeans2}) also $\rho\sigRs$ and $\rho(\tvphism-\sigRs)$ can
be written in the same way (their explicit expression is given in
Appendix A).

To split $\tvphism$ into streaming motion $\vcphi\equiv\tvphim$ (that for
simplicity we assume nowhere negative), and azimuthal dispersion,
$\sigphis\equiv\overline{(\tvphi -\vcphi)^2}=\tvphism -\vcphi^2$, we adopt
the Satoh (1980) $k$--decomposition:
\begin{equation}
\vcphi^2 = k^2(\tvphism -\sigRs),
\label{eq:vphi}
\end{equation}
and
\begin{equation}
\sigphis =\sigRs + (1-k^2) (\tvphism - \sigRs),
\label{eq:sigphi}
\end{equation}
with $0\le k\le 1$.  For $k=0$ no ordered motions are present, and the
velocity dispersion tensor is maximally tangentially anisotropic,
while for $k=1$ the velocity dispersion tensor is isotropic, and the
galaxy flattening is due to azimuthal streaming velocity (the so
called ``isotropic rotator''). In principle, by relaxing the
hypothesis of a constant $k$ and allowing for $k=k(R,z)$, even more
rotationally supported models can be constructed, up to the {\it
maximum rotation} case considered in Ciotti \& Pellegrini (1996),
where $k(R,z)$ is defined so that $\sigphis=0$
everywhere\footnote{Note that the important issue of the models
phase--space consistency is beyond the tasks of this work.}.

\subsection{Projected quantities}

To project the galaxy models on the plane of the sky (the projection
plane), we employ a Cartesian coordinate system $(\xp, \yp, \zp)$,
with the line of sight ({\it los}) directed along the $\zp$ axis, and with
the $\xp$ axis coincident with the $x$ axis of the natural Cartesian
system introduced at the beginning of Section 2.1.  The angle between
$z$ and $\zp$ is $\theta$, with $0\le\theta\le\pi /2$: $\theta=0$
corresponds to the face--on view of the galaxy, while $\theta=\pi/2$
to the edge--on view.  With this choice, the projection plane is
$(\xp,\yp)$, and the \los direction in the natural coordinate system
is given by $\nv =(0,-\sin\theta ,\cos\theta)$\footnote{The \los
vector points {\it toward} the observer, and so {\it positive}
velocities correspond to a {\it blue--shift}.}.  Accordingly, the
coordinates of the two Cartesian systems are related by
\begin{equation}
\cases{
x = \xp ,\cr
y = \yp\cos\theta -\zp\sin\theta ,\cr
z = \yp\sin\theta +\zp\cos\theta ,\cr}
\label{eq:xxp}
\end{equation}
and the homeoid labeled by $m$ can be rewritten in the observer
coordinate system as:
\begin{equation}
m^2=\left(\sqrt{A}\zpt +{B\over \sqrt{A}}\ypt\right)^2 +\ltild^2,
\label{eq:m2}
\end{equation}
where from now on, the symbol ``$\sim$'' over a coordinate will
indicate normalization to $\Rt$, and
\begin{equation}
\cases{
A\equiv\sin^2\theta +\cos^2\theta /q^2, \cr
B\equiv (1/q^2 -1)\sin\theta\cos\theta , \cr
\ell^2\equiv\xp^2 +\yp^2/\qt^2, \cr
\qt^2\equiv\cos^2\theta +q^2\sin^2\theta .\cr}
\label{eq:ABqt}
\end{equation}
When integrating a model quantity along the \los at given $(\xp,
\yp)$, the limits on $\zp$ are derived by setting $m=1$ in
eq. (\ref{eq:m2})
\begin{equation} 
\zpt_\pm =-{B\over A}\ypt\pm\sqrt{{1-\ltild^2\over A}}. 
\label{eq:zp}
\end{equation}
For examples, the surface density profile is given by
\begin{displaymath}
\Sigtl \equiv \int_{\zpm}^{\zpp}\rho\, d\zp =
\end{displaymath}
\begin{equation}
= \rhoc\,\Rt{q\over \qt}\betaf\left({1\over 2},n+1\right)
(1-\ltild^2)^{n+1/2},
\label{eq:Sigtheta}
\end{equation}
where $\rho(\xp ,\yp ,\zp)$ is obtained by substitution
eq. (\ref{eq:m2}) in eq. (\ref{eq:rho}).  The quantity $\ell$
determines the size of the elliptic isophotes, and their (constant)
{\it apparent ellipticity} is $\varepsilon =1-\qt$.  For a fixed
$\ell$, the major and minor isophotal semiaxes are $a=\ell$ and
$b=\qt\ell$, and the associated {\it circularized radius} is defined
by the identity $\pi R_{\ell}^2=\pi ab$, i.e.,
$R_{\ell}=\sqrt{\qt}\ell$.  In particular, the {\it circularized
effective radius} $\Re$ is given by
\begin{equation}
\Re =\sqrt{\qt}\ellef, 
\label{eq:recirc}
\end{equation}
where $\ellef$ is the solution of the equation $\Mt (\ellef)=M/2$, and
where the projected mass within $\ell$ is
\begin{displaymath}
\Mtl\equiv\int_{\ell^{'}\leq\ell}\Sigtlp\, d\xp\, d\yp  = 
\end{displaymath}
\begin{equation}
=\rhoc\Rt^3 \,2\,\pi\, q\,\betaf\left({3\over 2},n+1,\right)
      \left[1-(1-\ltild^2)^{n+3/2} \right].
\label{eq:projmass}
\end{equation}
We obtain
\begin{equation}
\ellef =\sqrt{1-2^{-1/(n+3/2)}}\Rt\equiv\Reo ,
\label{eq:elleff}
\end{equation}
with $\Reo$ the effective radius of the model when seen face--on (or
in case of spherical symmetry). As can be easily proved, {\it the
identity $\Re =\sqrt{\qt}\Reo$ is a general property of all
axisymmetric homeoidal distributions, independently of their specific
density profile}.

To obtain the velocity fields at $(\xp,\yp)$ we integrate along the
\los their projected component on $\nv$.  This is done by
transforming the corresponding spatial velocity moments from
cylindrical to Cartesian coordinates (see Appendix B).  For example,
the \los component of the {\it streaming} velocity field is
\begin{equation}
\vn \equiv \overline{\langle\vv ,\nv\rangle} =\vci\nci,
\label{eq:vn_def}
\end{equation}
where ${\langle ,\rangle}$ is the standard inner product and the
repeated index convention has been applied.  The analogous quantity
associated to the velocity dispersion tensor is
\begin{equation}
\signs\equiv \overline{\langle \vv-\vec{v},\nv \rangle^2} =
\sigij\nci\ncj.
\label{eq:sign_def}
\end{equation}
By using the two definitions above, and eqs. (B.4)-(B.5), the
expressions for $\vn$ and $\signs$ are:
\begin{equation}
\vn=-\vcphi\cos\varphi\sin\theta,
\label{eq:vn}
\end{equation}
and
\begin{equation}
\signs =\sigRs +(1-k^2)(\tvphism-\sigRs)\cos^2\varphi\sin^2\theta ,
\label{eq:sign}
\end{equation}
where the last identity is obtained by using eq. (\ref{eq:sigphi}).
The corresponding (mass--weighted) projected fields are obtained by
changing coordinates in eqs. (\ref{eq:vn}) and (\ref{eq:sign}), and
then integrating on $\zp$:
\begin{equation}
\Sigtl\vp(\xp,\yp)\equiv\int_{\zpm}^{\zpp}\rho\vn d\zp ,
\label{eq:Sigvp}
\end{equation}
\begin{equation}
\Sigtl\Vps(\xp,\yp)\equiv\int_{\zpm}^{\zpp}\rho\vn^2 d\zp,
\label{eq:SigV2p}
\end{equation}
and
\begin{equation}
\Sigtl\sigps(\xp,\yp)\equiv\int_{\zpm}^{\zpp}\rho\signs d\zp.
\label{eq:Sigsigp}
\end{equation}
In general $\sigps$ will not coincide with the velocity dispersion we
measure in observations: in fact, in presence of a non--zero projected
velocity field $\vp$, the correct definition for this quantity is
\begin{displaymath}
\Sigtl\sigobs(\xp ,\yp)=\int_{\zpm}^{\zpp}\rho\,\,
\overline{\left(\langle\vv ,\nv\rangle - \vp\right)^2} d\zp =
\end{displaymath}
\begin{equation}
=\Sigtl\left(\sigps+\Vps-\vp^2\right),
\label{eq:Sigsiglos}
\end{equation}
where the last expression is derived from the identity
$\overline{\langle\vv ,\nv\rangle^2}=\signs +\vn^2$.  Note that,
independently of the \los orientation, on the isophotal minor axis
$\yp$ (where, by definition, $\cos\varphi =0$) $\vn$, $\Vps$, and
$\vp^2$ vanish, and $\sign=\sigR$: on this axis $\sigps$ is the
projection of $\sigRs$ and $\sigobs =\sigps$. In addition, the last
identity holds everywhere when observing the galaxy face--on ($\theta
=0$), or in the case $k=0$.  Since the observed velocity dispersion is
always measured within a given aperture, we finally integrate
$\sigob^2$ over the isophotes (even though $\sigob$ in general is not
constant over isophotes):
\begin{equation}
\Mtl\sigobsal\! =\! \langle\Sigt\sigobs \rangle_\ell\!\equiv\!\!\!
\int_{\ell^{'}\leq\ell}\!\!\!\!\!\!\!\!\Sigtlp\sigobs(\xp,\yp)\,\!d\xp\, d\yp .
\label{eq:sigobsa}
\end{equation}
In Appendix B we obtain the explicit expressions for $\sigps$, $\Vps$
and their aperture values.  Unfortunately, $\vp$ cannot be cast in
algebraic form when $n>0$, and so we have to resort to numerical
integration of eq. (\ref{eq:Sigvp}) for its evaluation; the details
are given in Section 3.

%--------------------------------------------------------------------
\begin{figure}[htbp] 
\parbox{1cm}{\psfig{file=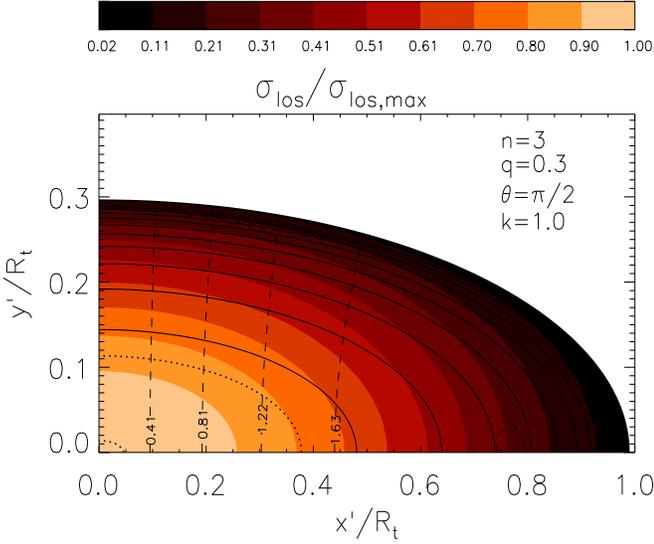,width=9.5cm,height=7.5cm,angle=0}}
\caption[]{The projected velocity fields of the $(n=3, q=0.3, k=1)$
model, when seen edge--on ($\theta=\pi/2$). The color contour plot
shows the velocity dispersion field $\sigob$ normalized to its maximum
value $\sigobm\simeq 0.23\,\Rt\,\sqrt{G\rhoc}$. Dashed lines represent
the projected rotation field $\vp/\sigobm$, with numerical values
labelled in the figure.  Solid lines are the surface brightness
isophotes $\mu= -2.5\log\Sigtl$ sampled at 1 magnitude difference,
while dotted lines are the isophotes corresponding to $\Re/8$ and
$\Re$. 
}
\label{fig:vpk1npi2}
\end{figure}
%-------------------------------------------------------------------

As a check of the exactness of the derived projected fields, we use a
general consequence of the projected virial theorem (see, e.g., Ciotti
2000), i.e.
\begin{displaymath}
\langle\Sigt\,(\sigps+\Vps)\rangle_{\ell=\Rt} =
2\,\nci\ncj\,K_{ij}=
\end{displaymath}
\begin{equation}
= -\ncu^2\Wuu -n_2^2 \Wdd -\nct^2\Wtt, 
\label{eq:projVT}
\end{equation}
where $K_{ij}$ is the kinetic energy tensor, and 
\begin{equation}
\Wij=-\int\rho\xci {\partial\phi\over\partial\xcj}\dtx,
\label{eq:Wij}
\end{equation}
are the components of the potential energy tensor.  For our models,
$\Wdd =\Wuu$, $\ncu=0$, and the explicit expressions of $\Wuu$ and
$\Wtt$ are given in Appendix A. 

We recall that a similar approach to the one presented in this Section
was adopted by vABS, who used a homeoidal, modified Jaffe density
profile; in particular, they studied the projected field
corresponding (in our notation) to the quantity $\sigobs +\vp^2=
\sigps+\Vps$ averaged within $0.5\Re$.

\section{The model properties}

To better understand the results of the Monte--Carlo simulations
presented in Section 4, here we illustrate in details a few
representative models, focusing on the effects of the various
parameters on the projected velocity fields, and on the observational
quantities entering the FP.

%--------------------------------------------------------------------
\begin{figure}[htbp] 
\parbox{1cm}{\psfig{file=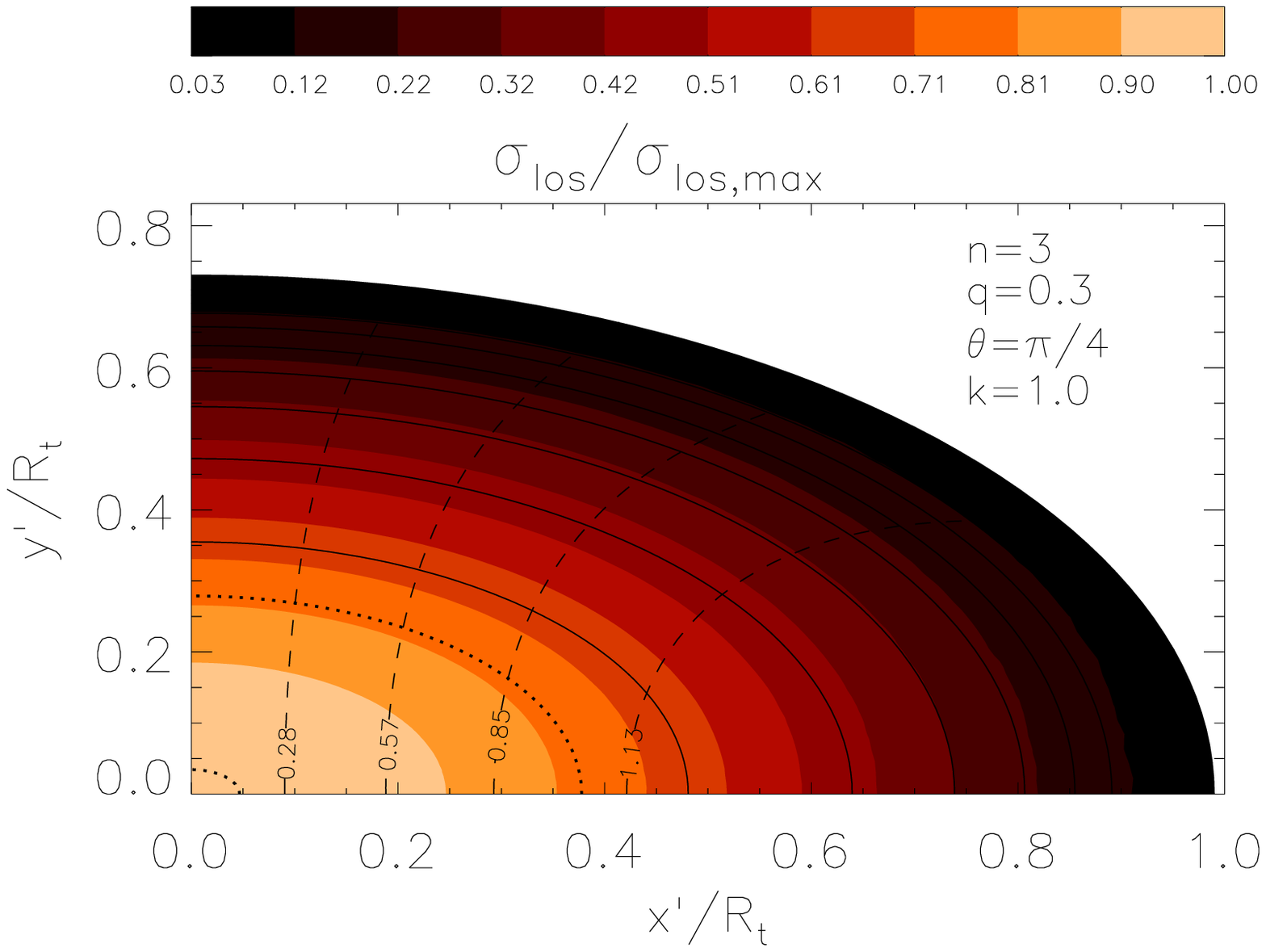,width=9.5cm,height=7.5cm,angle=0}}
\caption[]{The fields $\sigob/\sigobm$ and $\vp/\sigobm$ of the same
model in Fig. 1 when seen at $\theta =\pi/4$. In this case
$\sigobm\simeq 0.24\,\Rt\,\sqrt{G\rhoc}$.  }
\label{fig:vpk1npi4}
\end{figure}
%-------------------------------------------------------------------

The projected velocity dispersion and quadratic velocity fields in
eqs. (\ref{eq:Sigsigp}) and (\ref{eq:SigV2p}) are evaluated by using
the explicit expressions given in Appendix B. To obtain the projected
field $\vp$ in eq. (\ref{eq:Sigvp}) a numerical integration on $\zp$
is required. For symmetry properties we restrict the computation to
the first quadrant of the projection plane, that is organized with an
ellipsoidal grid made of 50 uniformly spaced isophotal contours.  Each
contour is divided in 50 angles, while the \los length $\zpp -\zpm$
(see eq. [\ref{eq:zp}]) is divided in 100 elements.  The computation
of the projected fields on this grid, by means of a double precision
Fortran90 code, requires $\simeq 20$ min on a $1.2$ GHz
workstation. To check the robustness and correctness of the code, for
all the explored models (many more than those presented) we have
verified the projected virial theorem given in eq. (\ref{eq:projVT}),
and we found relative errors $\lsim 10^{-3}$.

The illustrative cases presented here all refer to a model with $n=3$
and $q=0.3$, but their main properties apply to the whole family of
models studied in this paper.  Due to the constancy of the
mass--to--light ratio within each model, the mass weighted and the
luminosity weighted quantities are coincident.

%--------------------------------------------------------------------
\begin{figure}[htbp] 
\parbox{1cm}{\psfig{file=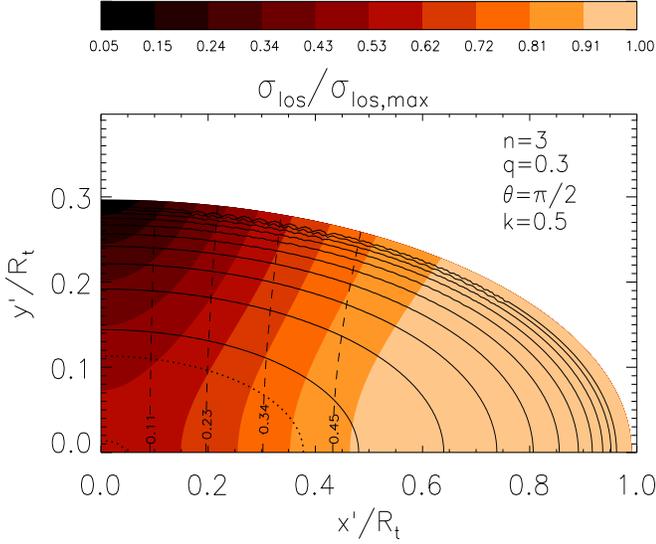,width=9.5cm,height=7.5cm,angle=0}} 
\caption[]{The fields $\sigob/\sigobm$ and $\vp/\sigobm$ of the model
in Fig. 1 when seen edge--on, in case of a substantial reduction of
the ordered motions ($k=0.5$). The maximum value of the velocity
dispersion field is $\sigobm \simeq 0.41\,\Rt\,\sqrt{G\rhoc}$; all the
comments in the caption of Fig. 1 apply.}
\label{fig:vpk05npi2}
\end{figure}
%-------------------------------------------------------------------

Figure 1 shows the edge--on view of the two observationally accessible
projected fields, $\sigob$ and $\vp$, in the isotropic rotator case.
As expected, the maximum value of $\sigob$ is reached at the center,
while $\sigob$ is not constant on isophotes. Note that {\it inside the
ellipse corresponding to a circularized radius of $\Re/8$} (an
aperture often used to correct $\sigc$ in the FP studies; e.g.,
JFK96), $\sigob$ {\it is constant well within 10\% (in fact, better
than 1\%)}.  As a consequence of the adopted decomposition of
azimuthal motions and of the edge--on view of the model, the projected
streaming velocity field $\vp$ (dashed lines) is nearly
vertical\footnote{The $n=0$ model is an exception: its lines of
constant $\vp$ are always parallel to the isophotal minor axis
(eq. [C.7]), and in the edge--on, isotropic rotator case, $\sigob$ is
constant on isophotes (eqs. [C.9]-[C.10]).}, its value decreases
towards the center of the galaxy, and vanishes on the $\yp$ axis.  As
anticipated in Section 2.1 the adopted density profiles, at variance
with real galaxies, are very flat in their inner regions: this is
clearly visible here, where the surface brightness
$\mu=-2.5\log\Sigma_\theta(\ell)$ drops from the center to $\Re$ by an
amount $\Delta \mu = [(2n+1)/(2n+3)]\,2.5\,\log 2\,\lsim \,1$, in
contrast with the drop of more than 8 magnitudes for $R^{1/4}$
galaxies.

In Fig. 2 we show the same model seen at $\theta=\pi/4$: for obvious
geometrical reasons, the lines of constant $\sigob$ are now more
similar to the optical isophotes. {\it The field $\sigob$ within
$\Re/8$ is still constant with very good approximation}, even though
$\Re$ has increased according to eq. (\ref{eq:recirc}).  The field
$\vp$ is deformed by projection effects, and its value, normalized to
the maximum of $\sigob$, is lower than in the edge--on case, as
expected.

When the amount of ordered rotation is substantially reduced (for
example, by assuming $k=0.5$), the resulting velocity fields are
modified as shown in Fig. 3 ($\theta =\pi/2$) and in Fig. 4 ($\theta
=\pi/4$).  Direct comparison with Figs. 1 and 2 indicates that such a
reduction of $k$ moves the maximum of $\sigob$ from the center to the
external regions of the model, a consequence of the increase of
$\sigphi$ at large galactocentric distances on the equatorial plane in
order to sustain the model flattening. This trend of $\sigob$ is
usually not observed in real galaxies, and it can be ascribed to the
too ``rigid'' Satoh decomposition: {\it however, $\sigob$ is again
nearly constant within $\Re/8$}.

%--------------------------------------------------------------------
\begin{figure}[htbp] 
\parbox{1cm}{\psfig{file=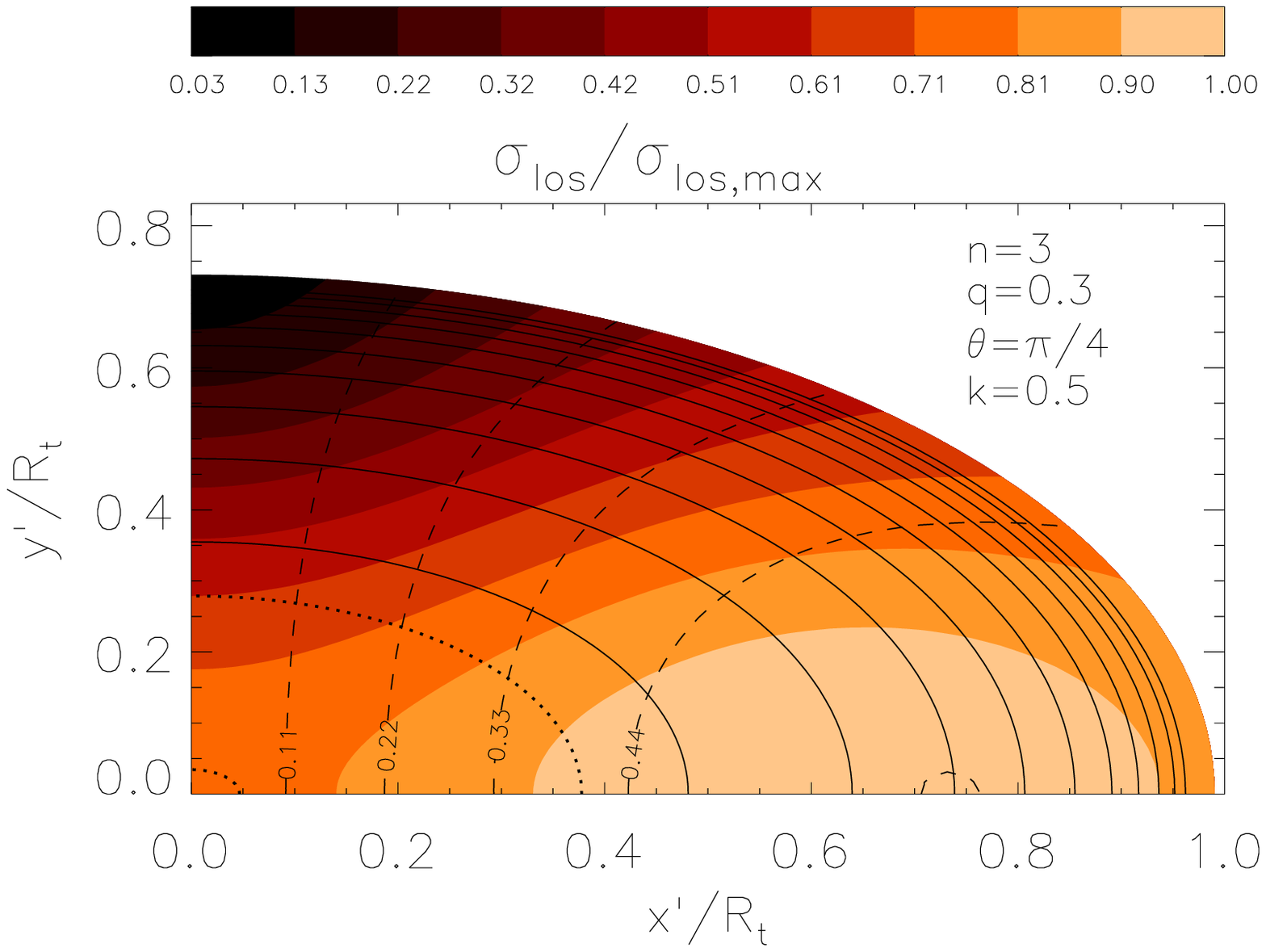,width=9.5cm,height=7.5cm,angle=0}}
\caption[]{The fields $\sigob/\sigobm$ and $\vp/\sigobm$ of the same
model in Fig. 3 when seen at $\theta =\pi/4$. In this case
$\sigobm\simeq 0.31\,\Rt\,\sqrt{G\rhoc}$.  }
\label{fig:vpk05npi4}
\end{figure}
%-------------------------------------------------------------------

While we have shown here a few illustrative examples, we find that
for all the models studied in detail $\sigob$ stays nearly constant
within $\Re/8$.  For example, for the $n=0$ model, by expanding
eq. (C.11) for $\ltild << 1$ we obtain
\begin{equation}
{\sigobsal\over\sigp^2(0,0)}\sim 1 +\!
\left [{3 (1-k^2)\sin^2 \theta\over 2}\!
       \left({\wu\over q^2\wt}-1\right ) -1 \right ]{\ltild^2\over 2}, 
\end{equation}
and so $\sigp(0,0)$ and $\sigoba(\Re/8)$ differ less than 0.1\%, while
for the $n=3$ models represented in Figs. 1-4 the two quantities
differ less than 1\%. This implies that when using apertures of the
order of $\Re/8$, the average $\los$ velocity dispersion can be safely
replaced by its central value, i.e.,
$\sigoba(\Re/8)\simeq\sigob(0,0)=\sigp(0,0)$, {\it independently of
rotation and \los inclination angle} (note that the last identity
holds exactly; see Section 2.2). These considerations are also
confirmed by the cuspier models discussed in vABS.

\begin{figure}[htbp]
\parbox{1cm}{\psfig{file=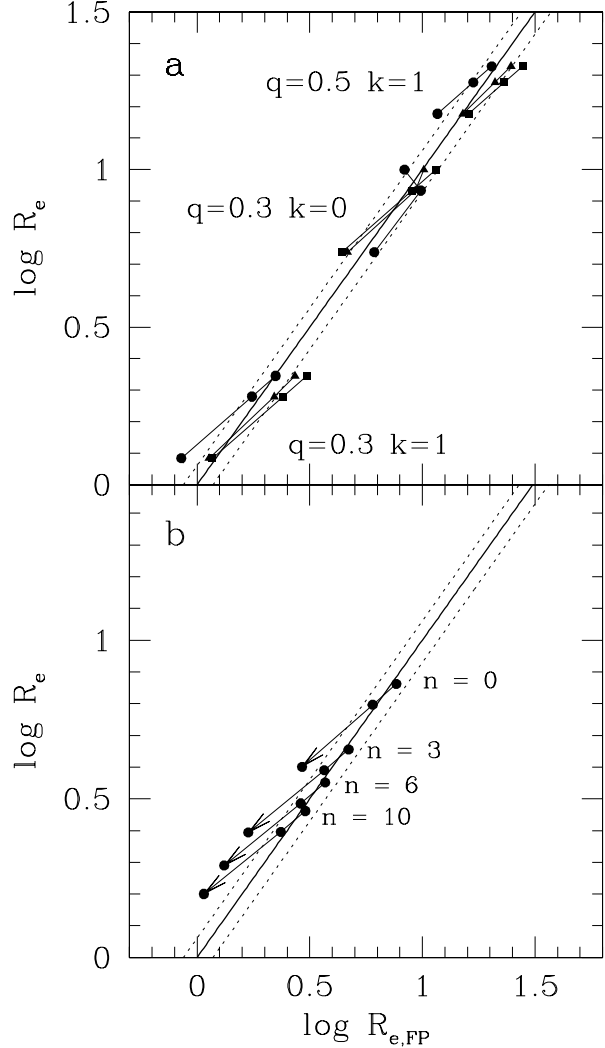,width=15cm,height=15cm,angle=0}}
\caption[]{{\it Panel a}: effects of \los direction and spectroscopic
aperture on three $n=0$ models, arbitrarily placed on the edge-on FP
(solid line), as a function of flattening ($q$) and amount of ordered
rotation ($k$). Dotted lines mark $\onesigint$. In each model, the
three aperture radii are $\Re/8$ (squares), $\Re$ (triangles), whole
aperture (circles), and for each aperture radius, the three points at
decreasing $y$-axis correspond to $\theta=0,\,\pi/4,\,\pi/2$. {\it
Panel b}: effects of \los direction and density profile on the
position of the models along the edge--on FP, when $\sigc
=\sigp(0,0)$. As in panel a, $\theta$ increases along the arrows.}
\label{fig:move_fp}
\end{figure}

We now describe how models ``move'' in the edge--on view of the FP,
when changing the intrinsic (i.e., $n$, $q$, and $k$) and
observational ($\theta$) parameters. In Fig. \ref{fig:move_fp}a we
illustrate the behavior of three $n=0$ models in the ($\log\Refp$,
$\log\Re$) space, where $\log\Refp\equiv\alpha\log\sigc +\beta\log\Ie
+\gamma$.  Two models have maximum intrinsic flattening ($q=0.3$), and
differ for the amount of rotation ($k=1$, and $k=0$); the third one is
a rounder ($q=0.5$) isotropic rotator. Owing to the particularly
simple expression of $\sigoba$ (eq. [C11]), we also investigate the
effect of adopting different apertures for the estimate of $\sigc$. In
all cases, varying the projection angle from 0 to $\pi/2$, makes $\Re$
to decrease according to eq. (\ref{eq:recirc}), thus producing the
vertical down--shift of the representative models; obviously, such a
shift is smaller for the rounder galaxy.  When $\Re$ decreases $\Ie$
increases, and galaxy models move along straight lines of constant
inclination with respect to the edge--on FP, independently of their
specific density distribution (see comments below
eq. [\ref{eq:elleff}]), and provided that $\sigc$ is only weakly
dependent on the \los inclination.  We find that even when $\sigc$ is
measured within apertures of up to the order of $\Re$, the effect on
the model displacement in the FP is marginal, independently of the
viewing angle and the amount of rotational support. The only relevant
case is when the total aperture is considered (circles) {\it and} the
galaxy flattening is supported by the azimuthal velocity dispersion
(these results can be qualitatively interpreted by using eq. [C.11]
with $\ltild =1$)\footnote{Note that the observed velocity dispersion
entering the FP relation is usually corrected to a circular aperture
with diameter $1.19\,h^{-1}$ kpc (e.g., J{\o}rgensen at al. 1999),
corresponding to a radial range $\sim 0.05\,\Re$--$\Re$ for $h=0.5$,
and for typical values of $\Re$.  In any case, the FP equations
derived by using $\Re/8$ or the fixed metric aperture are in mutual
good agreement (JFK96).}. In Fig. \ref{fig:move_fp}b, the effect of
the viewing angle for different values of the shape parameter $n$ is
illustrated.  At variance with Fig. \ref{fig:move_fp}a, all models
have the same flattening ($q=0.3$), mass, mass--to--light ratio, and
truncation radius. The amount of projection effects is quantitatively
similar for different values of $n$, being mainly due to the
dependence of $\Re$ on the viewing angle. In summary, since in all
cases the directions along which models move are not parallel to the
FP best fit line, {\it projection effects do contribute to the
observed FP scatter, with effects $\lsim 2\onesigint$}.

Interestingly, from Fig. \ref{fig:move_fp}b it is also evident that
the trend of $n$ along the FP is in agreement with what found
observationally when galaxy light profiles are fitted with the $\ser$
models (e.g., Caon et al. 1993; CLR; Graham \& Colless 1997; Prugniel
\& Simiens 1997; Ciotti \& Lanzoni 1997; BCD): in fact, in this latter
class, an {\it increase} of $m$ corresponds to the galaxy density
profile being radially {\it flatter} in the external regions, as
Ferrers models behave for {\it decreasing} $n$.  However, the amount
of this effect in Ferrers density profiles is substantially
smaller. This can be estimated by considering the behavior of $\Kvir =
G M/\Re\sigps(0,0)$: in the spherical limit, $4.93\,\lsim
\,\Kvir(n)\,\lsim \,6$ for $0\leq n\leq 10$, to be compared with the
range $7.96\,\gsim\,\Kvir(m)\,\gsim\, 1.75$ for $1\leq m \leq 10$ in
case of $\ser$ models (BCD).  Thus, while structural non--homology
alone, with constant $\ml$, is sufficient to reproduce the whole tilt
of the FP in the case of the $\ser$ models (CLR; Ciotti \& Lanzoni
1997; BCD), this is not true for the Ferrers ellipsoids. Also note
that, at least in the spherical limit, the values of $\Kvir(n)$ are
within the range spanned by the virial coefficient in the case of the
$\ser$ models. In particular, they are close to the value of 4.65 that
characterizes the de Vaucouleurs profile $R^{1/4}$.  Since $\Kvir$ is
the only model-dependent property that explicitly enters the FP
relation (through eq. [\ref{eq:mlkvir}]), this ensures that the class
of models we are using is suitable for our investigation. The behavior
of $\Kvir(n)$ for different flattenings and viewing angles is
summarized in Table \ref{tab:kvir}.

%--------------------------------------------------------------------
%-------------------------------------------------------------------

%___________________________________ Two column table (place early!)
\begin{table} 
\caption[]{The quantity $\Kvir$ for Ferrers ellipsoids as a function
           of $n$, $q$, and viewing angle $\theta$, when $\sigc =\sigp
           (0,0)$.}      \label{Tab1} 
\halign{\strut# & #\hfil &\quad\hfil
                  #\hfil &\quad\hfil
                  #\hfil &\quad\hfil
                  #\hfil &\quad\hfil
                  #\hfil &\quad\hfil
                  #\hfil &\quad\hfil
                  #\cr \noalign{\hrule \vskip 5 pt} &
%$\Kvir$ & $q=1$ & $q=0.3,~\theta=0$ & $q=0.3,~\theta=\pi/2$ \cr 
$\Kvir$ & $q=1$ & $q=0.3$ & $q=0.3$ \cr 
       & &       & $\theta=0$ & $\theta=\pi/2$ \cr 
\noalign{\vskip 5 pt \hrule \vskip 5 pt} 
& $n=0$  & 4.93 & 8.29 & 15.13 &\cr 

& $n=3$  & 5.78 & 8.61 & 17.36 &\cr 

& $n=6$  & 5.92 & 8.68 & 17.72 &\cr 

& $n=10$ & 6.00 & 8.71 & 17.89 &\cr 
\noalign{\vskip 5 pt \hrule \vskip 2 pt}\cr}
%\vspace{4cm}
\label{tab:kvir}
\end{table}
%--------------------------------------------------------------------

%--------------------------------------------------------------------
\begin{figure}[htbp]
\parbox{1cm}{\psfig{file=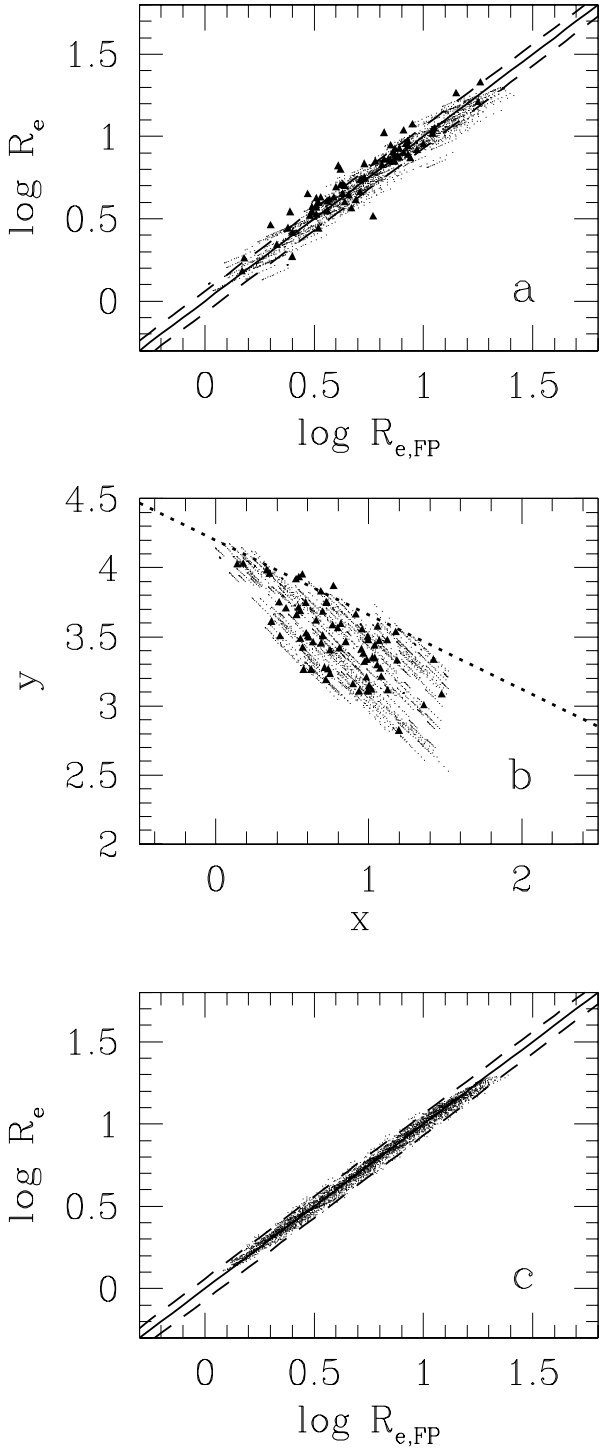,width=20cm,height=20cm,angle=0}}
\caption[]{Accepted models (dots) selected according to
eq. (\ref{eq:gauss}) and $\onesiginp=0.054$, plotted in the edge-on
({\it Panel a}), and face-on ({\it Panel b}) views of the FP.  For
models in Panel a, $\onesigint=0.057$.  Solid triangles are the Coma
cluster galaxies in Gunn r (from JFK96). {\it Panel c}: the edge-on FP
for the models selected in the ``zero--thickness''
approximation. Solid and dashed lines in Panels a and c represent
eq. (\ref{eq:FP}) and its intrinsic scatter, respectively. The dotted
line in Panel b marks the ``zone of exclusion'' (JFK96).
}
\label{fig:FP} 
\end{figure}
%-------------------------------------------------------------------

\section{Simulations and results}

\subsection{The numerical procedure}

In this Section we extend the statistical approach presented in BCD,
and we determine the most general manifold (in the parameter space)
defined by the models that lie on the observed FP.  In practice, for
each seven--dimensional point $(n,\ml,L,\Rt ,q, k, \theta)$ in the
model space, we determine $\Re$, $\Ie$, and $\sigc$.  Then, we check
if the model ``belongs'' to the FP.
The observational FP that we take as a reference is the one obtained
by JFK96 for the Coma cluster galaxies in Gunn r (eq. [\ref{eq:FP}]).
For its {\it intrinsic} scatter we adopt the value $\onesigint
= 0.057$, as quoted by JFK96 for the galaxies with $\sigc\ge 100$ km/s. 

The domains of model parameters considered in the simulations are the
following: $0\leq n\leq 6$, $1\leq \ml\leq 10$ (different from galaxy
to galaxy, but constant within each model), $2.7\le L\leq 50$ (in
$10^{10}\, L_\odot$, the same range of values spanned by the JFK96
Coma cluster galaxies), $1\leq\Rt\leq 200$ (in kpc), $0.3\le q\le 1$,
and $0\leq\theta\le\pi/2$.  The values of $n$ and $\ml$ are randomly
extracted from uniform distributions, while power-law distributions
$p(L)\propto L^{-1}$ and $p(\Rt)\propto \Rt^{-1.5}$ have been used to
extract $L$ and $\Rt$ by means of the von Neuman rejection technique.
The assumption of strongly non uniform input distributions for $L$ and
$\Rt$ was necessary in order to end (after the FP selection) with
galaxy models having a luminosity function and a distribution of
effective radii in agreement with those observed (see Sect. 4.3).  For
the extraction of the flattening $q$ a fit to the observed
distribution of intrinsic ellipticity (as derived for a population of
oblate spheroids by Binney \& de Vaucouleurs, 1981)
\begin{equation}
p(q)\simeq {0.6 (q+0.055)(q^2-1.07q+0.377)\over q^2-1.197q+0.372},
\label{eq:qdist}
\end{equation}
has been used. Even though the assumption of a population made of
oblate spheroids only is not fully consistent with observations (see,
e.g., Binney \& Merrifield 1998), it is acceptable for our
investigation, and it is also consistent with the geometry of the
adopted models.  Concerning the estimate of the model central velocity
dispersion, we recall that the way how models move along the FP is
almost the same when using apertures of $\Re/8$ or $\Re$, and so,
according to the results of Section 3, we assume $\sigc=\sigp(0,0)$.
In such a way the dimensionality of the parameter space is reduced by
excluding $k$ from the analysis.  Finally, to sample the effect of the
\los inclination, we compute the projections of each model along 11
viewing angles equally spaced in $0\le\cos\theta\le 1$.

For each projection angle we first check whether $\Re$ and $\sigc$ are
within the ranges $1\le\Re/{\rm kpc}\le 20$ and $100\le\sigc/({\rm
km/s)}\le 350$; if not, the model is discarded as unrealistic,
otherwise we construct the {\it angle average} quantities
$\langle\log\Re\rangle$, $\langle\log\sigc\rangle$,
$\langle\log\Ie\rangle$, and $\langle\log\Refp\rangle =
\alpha\,\langle\log\sigc\rangle +\beta\,\langle\log\Ie\rangle
+\gamma$, we calculate the quantity\footnote{We define residuals about
the FP the quantity $\dfp\equiv\log\Re- \log\Refp$}
$\langle\dfp\rangle \equiv \langle\log\Re\rangle
-\langle\log\Refp\rangle$, and we apply the following criteria to
check whether this candidate model ``belongs'' or not to the FP. Since
the FP residuals are consistent with being distributed as a Gaussian
(JFK96), we require that, according to the von Neuman rejection
technique, the {\it angle average model} is extracted from the
distribution
\begin{equation}
p(\langle\dfp\rangle)\propto\exp ({-\langle\dfp\rangle^2/2\,\onesiginp^2}),
\label{eq:gauss}
\end{equation}
where $\onesiginp$ is an input parameter. We finally accept the model
if it also belongs to the face-on FP, i.e., if it satisfies $0.54\,x +
y \lsim 4.2$, with $2.66\,x\equiv2.21\langle\log\Re\rangle
-0.82\langle\log\Ie\rangle+1.24\langle\log\sigc\rangle$, and
$1.49\,y\equiv1.24\langle\log\Ie\rangle+0.82\langle\log\sigc\rangle$
(JFK96).

The end product of a complete run is the data sample composed by the
11 projections of each accepted model. For this sample, in the
$(\log\Refp,\log\Re)$ space we estimate both the linear best--fit and
the $\rms$ of the residuals around the best fit line, $\rms(\dfp)$.
The procedure is repeated by changing the input parameters
$\onesiginp$ until $\rms(\dfp)=0.057$. The {\it physical} thickness of
the FP is then evaluated as $\rms(\langle\dfp\rangle)$, and the
contribution of projection effects is estimated as
$\onesigproj=\sqrt{\rms(\dfp)^2-\rms (\langle\dfp\rangle)^2}$.

%--------------------------------------------------------------------
\begin{figure*}[htbp] \centering
\includegraphics[width=\textwidth]{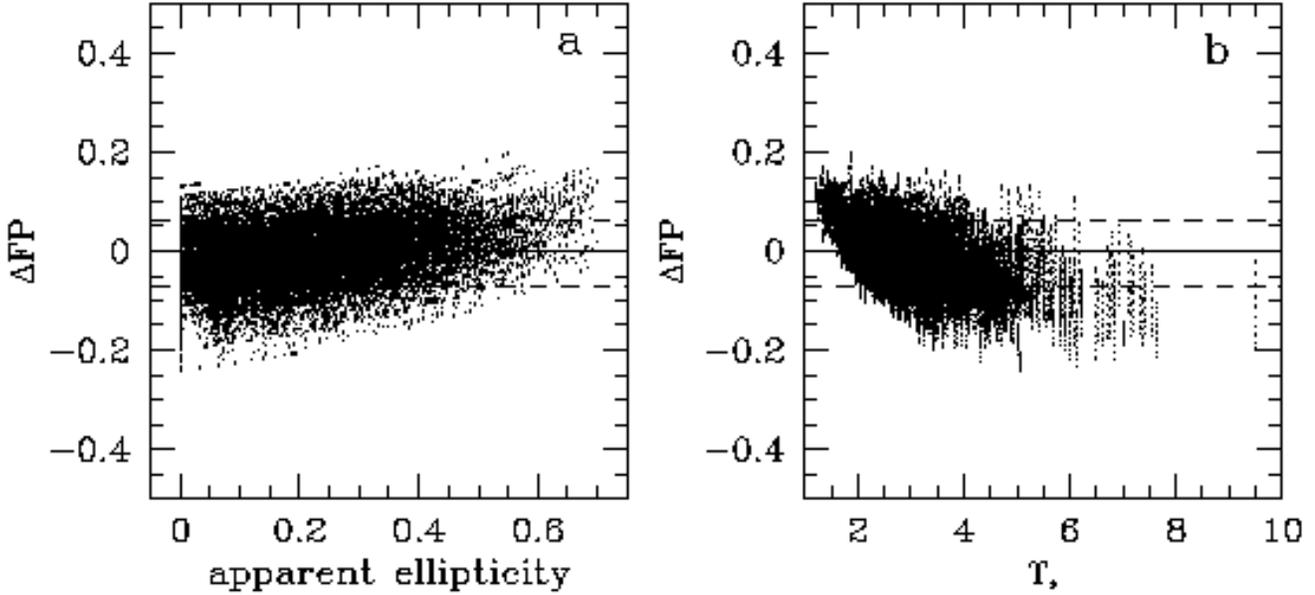}
\caption[]{Distribution of the residuals about the FP as a function of
apparent ellipticity (panel a) and mass--to--light ratio (panel b),
for models reproducing the observed FP. Points within the dashed lines
correspond to models with $\vert\Re/\Refp-1\vert\lsim 0.15$.  }
\label{fig:dfp}
\end{figure*}
%-------------------------------------------------------------------

Note that in this approach, due to the high dimensionality of the
parameter space, to the thinness of the FP, and to the von Neuman
rejections on $L$, $\Rt$, $q$, and $\langle\dfp\rangle$, we usually
need to calculate {\it several hundreds thousand projected models} for
each choice of $\onesiginp$.
This is feasible with the adopted class of models, because $\Re$ and
$\sigc$ can be expressed in a fully analytical way.
%For more realistic galaxy
%models, numerical solution and projection of the Jeans equations would
%be required, thus implying forbidding computational times.

\subsection{Projection effects on the FP thickness}

Following the procedure described above, we find that for $\onesiginp
\simeq 0.054$ the sample of accepted models defines a synthetic FP
that matches very well the observed one, both in the edge-on
(Fig. \ref{fig:FP}a) and in the face-on (Fig. \ref{fig:FP}b)
views. The model FP is characterized (by construction) by $\rms (\dfp)
=\onesigint$, while its {\it physical} thickness is $\rms
(\langle\dfp\rangle)\sim 0.052 =0.91\onesigint$. It follows that
$\onesigproj\sim 0.41\onesigint$. In this simulation, the fraction of
accepted models is $\sim 2.2\%$.

Another possibility to estimate the contribution of projection effects
to the FP thickness is that of selecting the angle averaged models
from what we call the ``zero--thickness'' FP: in practice, we adopted
$\onesiginp=0.001$ in eq. (34), so that the dispersion produced by the
accepted models when seen from the 11 different \los is {\it entirely}
due to projection. Note how, in agreement with the results shown in
Fig. 5, the final data set nicely fills the $1\sigma$ strip in
Fig. \ref{fig:FP}c. In these zero--thickness realizations, the
fraction of accepted model is $\sim 0.4\%$, and $\rms (\dfp)\simeq
0.45\,\onesigint$. Accordingly, we quantify the physical FP
scatter as $\onesigphys=\sqrt{\onesigint^2-\onesigproj^2}\simeq
0.89\onesigint$.

As a test of the robustness of the above estimates we also explored
the case in which the distribution of $\dfp$ is a step function, i.e.,
instead of using eq. (\ref{eq:gauss}), we accept the model if
\begin{equation}
\delta\equiv|10^{\langle\dfp\rangle}-1| \le\delinp,
\label{eq:muro}
\end{equation}
and we chose $\delinp$ so that $\rms(\dfp)$ of the selected models
equals $\onesigint$: the resulting values for $\onesigphys$ and
$\onesigproj$ are in perfect agreement with those obtained with
the previous approach.

For the whole sample of models that reproduces the observed FP no
systematic trends with the FP residuals are shown by $n$, $q$, $\Re$,
$L$, $\sigc$, while we find marginal correlations between the apparent
ellipticity $\epsilon=1-\qt$ and the mass-to-light ratio $\ml$, and
$\dfp$. In agreement with the analysis of JFK96 (cfr. Fig.8a therein)
and Saglia et al. (1993), flatter galaxies in projection are
preferentially characterized by positive residuals
(Fig. \ref{fig:dfp}a), while $\dfp$ decreases from positive to
negative values for increasing $\ml$ (Fig. \ref{fig:dfp}b).

\subsection{The FP tilt}

We now address the issue of the FP {\it tilt}, and compare the
properties of the models that reproduce the observed FP against the
available observational counterparts.

%--------------------------------------------------------------------
\begin{figure*}[htbp]
\centering
\includegraphics[width=\textwidth]{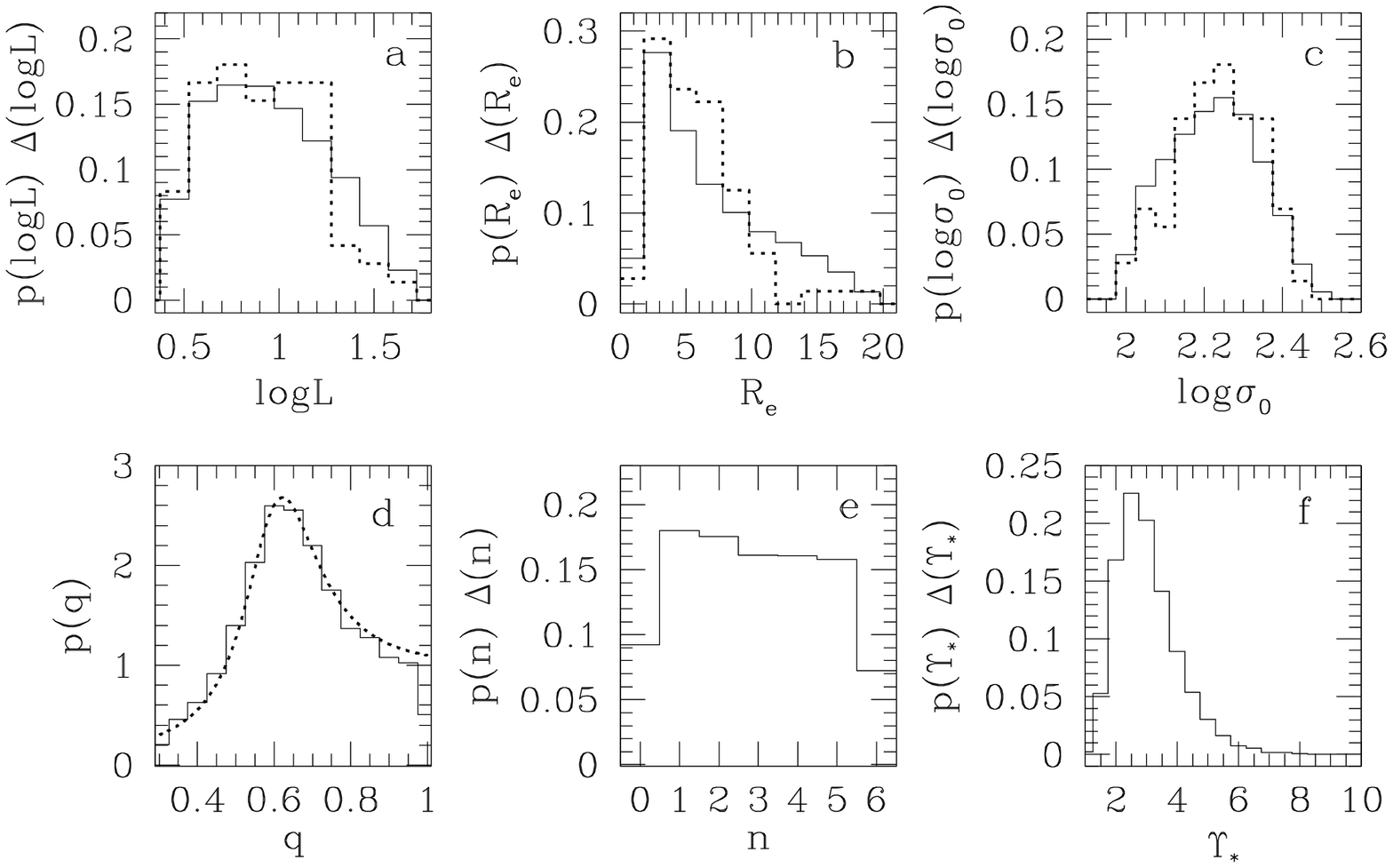}
\caption[]{Histograms of the properties of the models reproducing the
observed FP (solid lines). Dotted histograms in panels a, b, c
correspond to the observational data for the Coma cluster in Gunn r
(from JFK96).  The dotted line in panel d represents the input
distribution of $q$, as given in eq. (\ref{eq:qdist}).  }
\label{fig:histo}
\end{figure*}
%-------------------------------------------------------------------

In Fig. \ref{fig:histo} the distributions of the model properties are
shown with solid lines, while those of the adopted observational
sample are represented with dotted lines.  From Figs. \ref{fig:histo}
it is apparent how the FP selection modifies the input power-law
distributions of $L$ and $\Rt$ into distributions that match
remarkably well the observed ones for $L$ and $\Re$.  In particular,
the result should be contrasted with the one (not shown here) obtained
when extracting $L$ and $\Rt$ from {\it uniform} distributions: in
that case, $\log L$ and $\Re$ peak at $\sim$1.1 and $\sim$14,
respectively. An interesting case is presented by the flattening $q$:
in Fig. \ref{fig:histo}d it is apparent how the input 
distribution (dotted curve) is not modified by the FP selection.  Also
the effect on the shape parameter $n$ is not very strong, even if the
FP seems to be marginally selective against the lowest and the highest
values of $n$ (Fig. \ref{fig:histo}e).  On the contrary, the effect on
the mass-to-light ratio is remarkable: its input uniform distribution
has been substantially altered towards small values by the FP
selection (Fig. \ref{fig:histo}f).  Note the peak around $\ml\simeq
3$, a value in good agreement with the commonly accepted stellar
mass--to--light ratios in elliptical galaxies in the Gunn r band.

%--------------------------------------------------------------------
\begin{figure*}[htbp] \centering
\includegraphics[width=\textwidth]{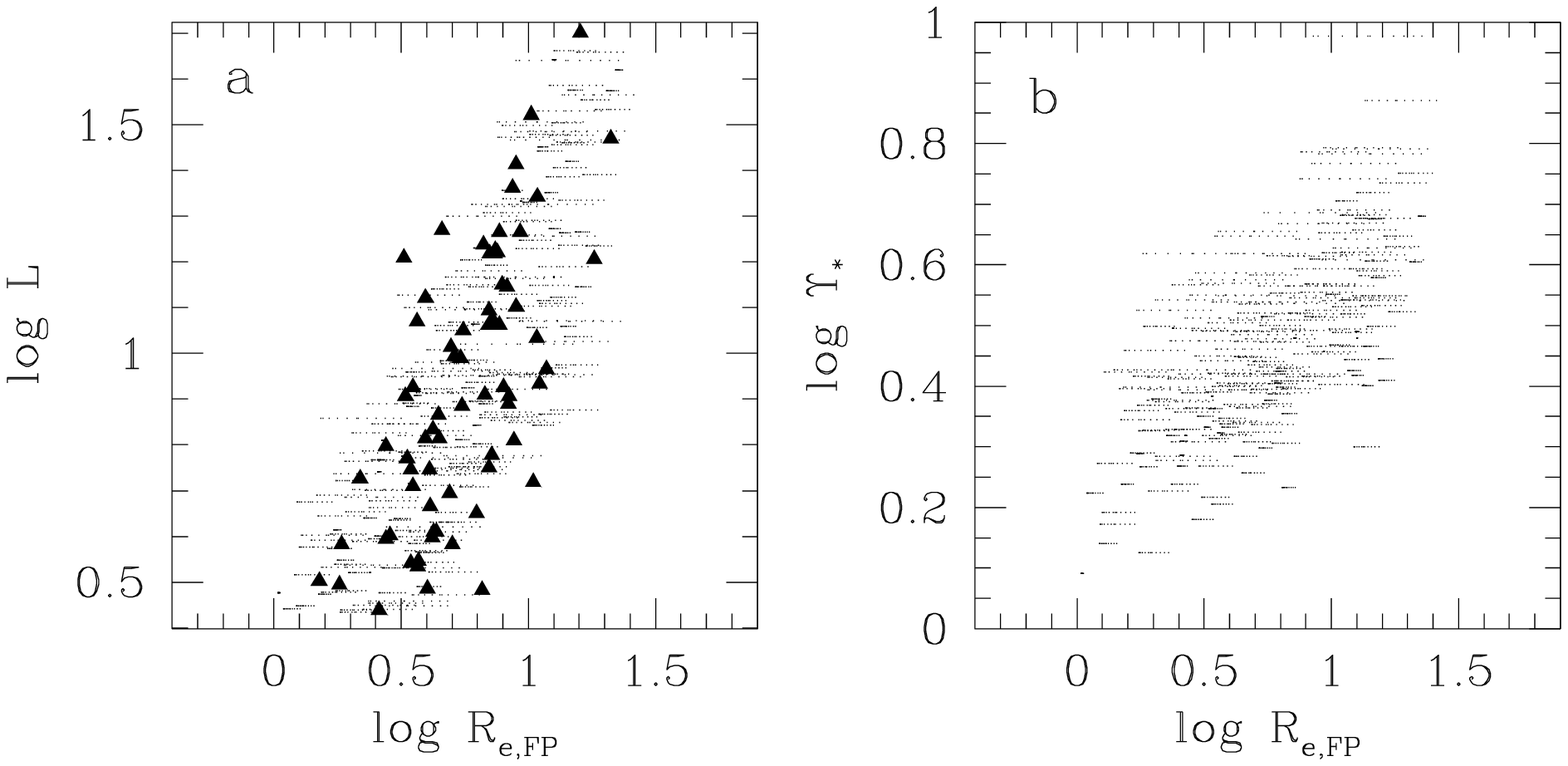}
%\begin{figure}[htbp]
%\parbox{1cm}{\psfig{file=vert_fig9.eps,width=16cm,height=16cm,angle=0}}
\caption[]{Distribution along the FP of luminosity and mass--to--light
ratio of the models that reproduce the observed FP (dots). Solid
triangles correspond to the observational data for the Coma cluster in
Gunn r (from JFK96).  }
\label{fig:tilt}
\end{figure*}
%-------------------------------------------------------------------

We now describe how the model properties vary along the FP.  While $n$
and $q$ do not show any particular trend or segregation, $L$ and $\ml$
are found to systematically increase with $\Refp$
(Fig. \ref{fig:tilt}).  In both cases, the scatter is large, but
almost constant for fixed $\Refp$.  As shown in Fig. \ref{fig:tilt}a,
not only the trend of $L$ with $\Refp$, but also the overall region
populated by the models in this space, correspond remarkably well to
those found in the observations.

\begin{figure}[htbp]
\parbox{1cm}{\psfig{file=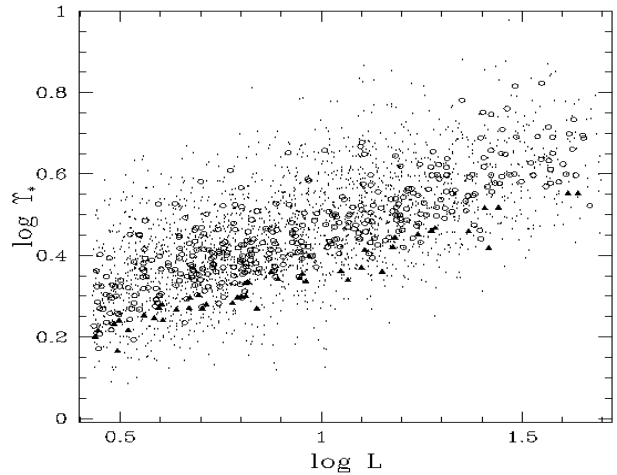,width=9cm,height=9cm,angle=0}}
\caption[]{$\ml$ vs. $L$ for the models that reproduce the observed FP
(small dots). Circles correspond to models belonging to the 
zero--thickness FP, while black filled triangles are their sub--sample
in which $0.9\le q\le 1$.
}  
\label{fig:lml} 
\end{figure}

The distributions of $L$ and $\ml$ along the FP translate in a well
defined mutual dependence of these model properties, as illustrated in
Fig. \ref{fig:lml} (small dots).  Also in this case the agreement
with the estimates from the observational data is remarkable
(cfr. Fig.3a in JFK96).  The large scatter in $\ml$ at fixed $L$
is by construction consistent with the small thickness of the FP:
apparently, other model properties vary within the sample of accepted
models so that their combined effect is to maintain the FP thin. A
clearer view of the situation can be obtained by considering only
galaxy models lying on the idealized zero--thickness FP (open
circles): the relation between $\ml$ and $L$ is better defined now,
even though the scatter in $\ml$ at fixed $L$ is still significant. In
case of an orthogonal exploration based on a systematic trend of $\ml$
with galaxy luminosity, the set of models would be just a
1--dimensional line in Fig. 10 (cfr. to Fig. 6 in BCD). Here, such a
case can be mimicked by restricting further to a sub--sample of models
characterized by a small range of flattenings (for example, $0.9\le
q\le1$, filled triangles in Fig. \ref{fig:lml}). For this latter data
set, the strict correlation $\ml\propto L^{0.3}$ is obtained, as
predicted by eq. (\ref{eq:mlkvir}) with $\Kvir\sim\,$const.

We recall that in this class of models $n$ has only a minor effect in
determining $\sigc$ (at variance with the shape parameter in the case
of $\ser$ models). However, the behavior of the density profiles and
$L$ is qualitatively the same in the two families, as discussed at the
end of Section 3.  In fact, when limiting to Ferrers ellipsoids with
constant $\ml$ and $q$, a correlation between the shape parameter and
the luminosity appears, with $n$ decreasing for increasing $L$.

%--------------------------------------------------------------------
%-------------------------------------------------------------------

\section{Discussion and Conclusions}

In this paper we have explored the importance of projection effects on
the FP thickness. We extended the statistical approach introduced by
BCD to a class of oblate galaxy models, with variable density profile,
flattening, amount of rotation, and fully analytical spatial and
projected dynamical properties. In particular, after generating galaxy
models corresponding to random choices of the model parameters, we
retained only those defining a FP with the same tilt and thickness as
the observed one.  With this approach not only we quantified the
importance of projection effects on the observed FP thickness (thus
also determining the ``physical'' FP scatter associated to the
dispersion in the galaxy internal properties), but we also studied in
a consistent way the possible origins of the FP tilt. Such a framework
represents a valuable alternative to the somewhat arbitrary orthogonal
explorations, where the property responsible for the FP tilt is
selected a priori, and a {\it fine tuning} problem for the selected
parameter is unavoidable. In the present approach instead, we find
that model properties can vary significantly from galaxy to galaxy,
while preserving the observed FP thinness (a result in agreement with
what found by BCD). Of course, we stress again that (at a deeper
level) the very existence of the FP with such a small scatter {\it is}
a fine tuning case, in the sense that stellar population ($\ml$) and
structural and dynamical properties ($\Kvir$) are tightly correlated,
as described by eq. (\ref{eq:mlkvir}). The reason for that can only be
found in a comprehensive theory of galaxy formation and evolution.

The main results of the present work can be summarized as follow:
\begin{itemize}

\item For the adopted class of models the contribution of ordered
streaming motions to the observed velocity dispersion is negligible
when small/medium apertures ($\lsim\,\Re$) are used for the
spectroscopic observations. This implies that a systematic decrease of
rotational support with increasing luminosity {\it is not} at the
origin of the FP tilt of elliptical galaxies.  However, the
contribution of rotation can be non negligible when using larger
apertures, and this must be taken properly into account when studying
the FP at high redshift.

\item When observed from different \los inclinations models move along
a direction that is not exactly parallel to the best fit line of the
edge--on FP, thus confirming that projection effects do contribute to
the observed FP scatter.  The amount of such a shift depends mainly on
the intrinsic flattening of the models (being larger for more
flattened systems) and, to a minor extent, on the adopted aperture
used for the determination of $\sigc$.

\item The estimated contribution of projection effects to the observed
dispersion of $\log\Re$ around the FP is $\sim 0.4\,\onesigint$, while
the FP {\it physical scatter} (as determined by variations of the
physical properties from galaxy to galaxy), is $\sim
0.9\,\onesigint$. It follows that, when studying the correlations
between galaxy properties required to reproduce the FP tilt and
thickness, spherical models are an acceptable approximation.

\item Weak correlations of the FP residuals with galaxy
mass--to--light ratio and apparent ellipticity are found.  The latter
is in good agreement with the results of JFK96 and Saglia et
al. (1993).

\item For the models that reproduce the observed FP, a very good
agreement between their properties ($L$, $\Re$, $\sigc$, and $q$) and
the observational data if found. Also the trend of $\ml$ with $L$
corresponds very well to the one estimated from the observations.  In
order to get these results, it has been crucial to extract the models
from steep power-law distributions of $L$ and $\Rt$, while {\it
uniform} distributions in input produce an excess of models with
bright luminosities and large effective radii. While the conclusions
about the contribution of projection effects to the physical FP
thickness would remain unchanged, it is still unclear why the
requirement that models belong to the FP is so selective against low
$L$ and $\Re$.
% In addition, it interesting is that by
% imposing to the models to also populate the face-on FP as observed,
% their velocity dispersion distribution matches very well the
% observational data (while without this further requirement, an excess
% of high velocity dispersion models would be found).

\item For what concerns the FP tilt, $L$ and $\ml$ of the accepted
models appear to systematically increase with $\Refp$. The
corresponding increase of $\ml$ with $L$ is also well defined and in
very good agreement with the observational estimates. In addition, in
the ($\ml,\,L$) plane, the models appear to be segregated in terms of
the flattening: at any fixed $L$, systems with low $\ml$ are
preferentially rounder than those with high $\ml$.

\item The ranges of variation of the shape parameter $n$, flattening,
and mass--to--light ratio for models of given luminosity can be very
large, although consistent (by construction) with a thin FP.  This
naturally solves the fine tuning problem met by the ``orthogonal
exploration'' approach, and it provides better agreement with
observational data.  In practice, model parameters mutually combine in
such a way that a large dispersion of galaxy properties is allowed at
any fixed location of the FP, while preserving its thinness.

\end{itemize}

For simplicity we have used, as a guiding tool for the present
investigation, simple one--component, two--integrals oblate galaxy
models.  The study would be best carried out with other families of
models, better justified from the observational and physical point of
view, but for the present purposes the simple models used provide an
adequate demonstration.  In fact, although the density distributions
adopted here are not a good representation of real elliptical
galaxies, the conclusions obtained about the projection effects on the
FP can be considered robust. In fact, it is also reassuring that the
displacements in the FP space of our models due to a change in the
\los direction agree both qualitatively and quantitatively with the
analogous results obtained from the end--products of N--body
simulations.  This is because the way $\Re$ (and thus $\Ie$) depends
on the viewing angle is the same for all homeoidally stratified
density distributions. A substantial improvement of the present
exploration, assuming more realistic density profiles, would require
much more time expensive simulations, since numerical calculation of
the projected velocity dispersion would be necessary.
\begin{acknowledgements}

We would like to thank Giuseppe Bertin, Gianni Zamorani and the
anonymous referee for very useful comments that greatly improved the
manuscript, and Reinaldo de Carvalho, Renzo Sancisi and Roberto Saglia
for useful discussions. L.C. also acknowledges the warm hospitality of
Princeton University Observatory, where a substantial part of this
work has been done. This work has been supported by MURST of Italy
(Co-fin 2000).

\end{acknowledgements}

\appendix
\section{3D quantities}

\subsection{Potential expansion and gravitational energy}

The general quadrature formula for the potential at internal points of
heterogeneous homeoidal distributions, can be rewritten in the
specific case of oblate Ferrers ellipsoids as
\begin{equation}
\phi(R,z) = -\pi G\Rt^3 q
            \int_0^\infty {\Delta\Psi(\tau)\over\Delta(\tau)}d\tau,
\end{equation}
where
\begin{equation}
\Delta\Psi(\tau)\equiv 2\int_{m(\tau)}^\infty\rho(t)tdt=
                \rhoc {[1-m^2(\tau)]^{n+1}\over n+1}, 
\end{equation}
\begin{equation}
m^2(\tau)\equiv {R^2\over \Rt^2+\tau}+{z^2\over q^2\Rt^2+\tau},
\end{equation}
and
\begin{equation}
\Delta(\tau) \equiv (\Rt^2 +\tau)\sqrt{q^2\Rt^2+\tau}.
\end{equation}
After a double expansion of identity (A.2) a cumbersome but trivial 
algebra shows that eq. (A.1) can be written as 
\begin{equation}
\phi(R,z) =-\pi G\rhoc\Rt^2\sum_{i=0}^{n+1}\sum_{j=0}^i
\phinijq\Rtild^{2(i-j)}\ztild^{2j},
\end{equation}
where $\Rtild\equiv R/\Rt$, $\ztild\equiv z/\Rt$,
\begin{displaymath}
\phinijq\equiv {q (-1)^i\over (n+1)(i+1/2)} 
               {n+1\choose i}{i\choose j}\,
\end{displaymath}
\begin{equation}
~~~~~~~~~~~~~~ \hypgeof\left (j+{1\over 2}, i+{1\over 2}, i+{3\over 2};
1-q^2\right), 
\end{equation}
and $\hypgeof$ is the standard hypergeometrical function defined by
\begin{equation}
\hypgeof(a,b,c;z)\!=\!{\Gamma(c)\over \Gamma(b)\Gamma(c-b)}\!
                  \int_0^1 \!\! t^{b-1}(1-t)^{c-b-1}(1-zt)^{-a}dt, 
\end{equation}
with $c>b>0$ and $\mid z\mid < 1$.  Note that $\hypgeof(a,b,c;0)=0$
and that for $i$ and $j$ integers, $\hypgeof$ in eq. (A.6) reduces to a
combination of elementary functions.

The components of the gravitational energy tensor used at the end of
Subsection 2.2 can be evaluated from the useful identities
\begin{equation}
\Wij = -G\Rt^4\pi^2 q a^2_i w_i \delta_{ij}
\int_0^\infty \Delta\Psi^2(m) dm,
\end{equation}
where $a_1,\,a_2,\,a_3$ are the three semi-axes of the homeoid, $w_i$
are functions of the flattening $q$, $\Delta\Psi$ is given in
eq. (A.2), and $\delta_{ij}$ is the Kronecker index (Roberts 1962).
For the models considered in this work $\Wdd = \Wuu$, and the explicit
expressions of $\Wuu$ and $\Wtt$ are:
\begin{equation}
\Wuu = -G \rhoc^2\Rt^5{\pi^2 q\wu\over 2(n+1)^2} 
\betaf\left({1\over 2},2n+3\right), 
\end{equation}
and
\begin{equation}
\Wtt =-G\rhoc^2\Rt^5{\pi^2 q^3\wt\over 2(n+1)^2}
\betaf\left({1\over 2},2n+3\right).
\end{equation}
Note that in the notation of eq. (A.5) $\wu=\phi_{010}(q)$ and
$\wt=\phi_{011}(q)$, i.e.,
\begin{equation}
\wu ={q\over 1-q^2}
\left({\arcsin\sqrt{1-q^2}\over \sqrt{1-q^2}} -q\right),
\end{equation}
and
\begin{equation}
\wt = {2q\over 1-q^2}
\left({1\over q} - {\arcsin\sqrt{1-q^2}\over \sqrt{1-q^2}}\right),
\end{equation}
(see, e.g., BT). In the limit of nearly round models ($q\to 1$)
$\wu\sim 2(1+2q/3)/5 + {\rm O}(1-q)^2$ and $\wt\sim 2(3-4q/3)/5 + {\rm
O}(1-q)^2$.

\subsection{Velocity dispersions expansion}

We start by noting that eq. (\ref{eq:rho}) with $n$ integer can be
rewritten as
\begin{equation}
\rho(m) =\rhoc\sum_{i=0}^n\sum_{j=0}^i\rhonijq\, \Rtild^{2j}\,
\ztild^{2(i-j)}, \quad (m\leq 1),
\end{equation}
where
\begin{equation}
\rhonijq\equiv {(-1)^i\over q^{2(i-j)}}{n\choose i}{i\choose j}. 
\end{equation}
With the aid of this expansion, we can proceed to the solution of
eqs. (\ref{eq:soljeans1})-(\ref{eq:soljeans2}): at first sight the
simplest way could appear the direct substitution in these equations
of expansions (A.5) and (A.13).  On the contrary, even though these
substitutions show that the solution of the Jeans equations can be
obtained as finite expansions, for computational reasons this is not
the most efficient way to perform the calculations.  For example, by
using the fact that $\rho$ and $\phi$ depend only on $z^2$, in
eq. (\ref{eq:soljeans1}) we changed integration variable from $z$ to
$z^2$, we expanded the quantity $\partial\Delta\Psi (\tau)/\partial
z^2$ in eq. (A.1), and finally we integrated the resulting expression
multiplied by eq. (A.13). The result is
\begin{equation}
\rho\sigRs =G\Rt^2\rhoc^2\!
            \sum_{i=0}^n\sum_{j=0}^i\sum_{k=0}^a\sum_{l=0}^k 
            \zetanijklq\Rtild^{2(j+l)}\ztild^{2(k-l)},
\end{equation}
where $a=2n+1-i$, and $\zetanijklq\equiv\zeta_1\zeta_2\zeta_3$ with 
\begin{equation}
\zeta_1 =\pi\, q^{2(n-i-k+l)+3}\,(1-q^2)^j,
\end{equation}
\begin{equation}
\zeta_2 ={(-1)^{j+k}\over 2n+1-i}
         {n\choose i}{i\choose j}{2n+1-i\choose k}{k\choose l}, 
\end{equation}
\begin{displaymath}
\zeta_3 =\betaf\left(n+{3\over 2}-i+j,i+1\right)\,
\end{displaymath}
\begin{equation}
~~~~~~~~~ \hypgeof\left(n+{3\over 2},n+{3\over 2}-i+j,n+{5\over
2}+j;1-q^2\right). 
\end{equation}
The r.h.s. of eq. (\ref{eq:vphi}) is the sum of two contributions. The
expansion of $R\partial\rho\sigRs/\partial R$ is immediately obtained
from eq. (A.15), while following the same approach used for the
derivation of eq. (A.15) it can be proved that
\begin{equation}
R\,\rho\,{\partial\phi\over\partial R}=G\Rt^2\rhoc^2
\sum_{i=0}^n \sum_{j=0}^i \sum_{k=0}^n \sum_{l=0}^k
\etanijklq\,\Rtild^{2a}\,\ztild^{2b}, 
\end{equation}
where $a=j+l+1$, $b=i-j+k-l$, and 
\begin{displaymath}
\etanijklq\equiv {2\pi\over q^{2(k-l)-1}}{(-1)^{i+k}\over i+3/2}
{n\choose i}{i\choose j}{n\choose k}{k\choose l}\,
\end{displaymath}
\begin{equation}
~~~~~~~~~~~~~~~~ \hypgeof\left(i+{1\over 2}-j,i+{3\over 2},i+{5\over 2};
                 1-q^2\right). 
\end{equation}
Note that a simple check of eqs. (A.15) and (A.19) shows that the
cylindrical radius $R^2$ can always be factorized in the expression of
the quantity $\rho (\tvphism-\sigRs)$: in other words, the latter
quantity vanishes (as expected) on the symmetry axis of the model.

\section{Projected quantities}

\subsection{From cylindrical to Cartesian coordinates}

To project the velocity fields on the plane of the sky according to
eqs. (\ref{eq:Sigvp})-(\ref{eq:sigobsa}), we first transform
velocities from cylindrical coordinates, where the Jeans equations are
solved, to the natural cartesian coordinates, in which the projection
is particularly simple (eqs. [\ref{eq:vn_def}]-[\ref{eq:sign_def}]) 
\begin{equation}
\cases{
\tvx = \tvR\cos\varphi -\tvphi\sin\varphi, \cr
\tvy = \tvR\sin\varphi +\tvphi\cos\varphi, \cr
\tvz = \tvz.
}
\end{equation}
By definition
\begin{equation}
\vci \equiv \overline{\ttv}_i ={1\over \rho}\int f(\xv ,\vv ,t)\dtvv ,
\end{equation}
and 
\begin{equation}
\sigij\equiv \overline{(\ttv_i -\vci)(\ttv_j-\vcj)} =
\overline{\ttv_i\ttv_j}-\vci\vcj,
\end{equation}
and so, following the choice of $f=f(E,\Lz)$, from eq. (B.1)
\begin{equation}
\cases{
\vcx =-\vcphi\sin\varphi, \cr
\vcy = \vcphi\cos\varphi, \cr
\vcz = 0,}
\end{equation}
and
\begin{equation}
\cases{
\sigxx= \sigRs\cos^2\varphi +\sigphis\sin^2\varphi, \cr
\sigyy= \sigRs\sin^2\varphi +\sigphis\cos^2\varphi, \cr
\sigzz= \sigRs, \cr
\sigma_{xy} = (\sigRs -\sigphis)\sin\varphi\cos\varphi, \cr
\sigma_{xz} = \sigma_{yz} = 0.}
\end{equation}

\subsection{Projection of $R^{2a}z^{2b}$}

In order to integrate along the \los eqs. (\ref{eq:Sigvp}) and
(\ref{eq:Sigsigp}), we first note that $\cos\varphi =x/R=\xp/R$, i.e.,
the term $\cos^2\varphi$ brings in the integrand only the quantity
$R^{-2}$, being $\xp$ fixed at the chosen position in the projection
plane.  In addition, from eqs. (\ref{eq:soljeans2}), (A.15), and (A.19),
it results that $\rho(\tvphism -\sigRs)$ is a sum of terms
$\Rtild^{2a}\ztild^{2b}$, with $a$ and $b$ non negative integers.  As
a consequence, the projection of the quadratic velocity fields reduces
to the evaluation of finite sums of terms of general form:
\begin{equation}
I_{a,b}(\xp, \yp) \equiv\int_{\zpm}^{\zpp}\!
                          \Rtild^{2a}\,(\xpt ,\ypt, \zpt)\,
                          \ztild^{2b}(\xpt ,\ypt ,\zpt)\,d\zp ,  
\end{equation}
where $R$ and $z$ are expressed in terms of $\xp$, $\yp$, and $\zp$
following eq. (\ref{eq:xxp}). The result is
\begin{equation}
I_{a,b}(\xp ,\yp) =\!\Rt \!
                   \sum_{\alpha=0}^a 
                   \sum_{\beta=0}^{\alpha}
                   \sum_{\gamma=0}^{2b} 
                   \sum_{\delta=0}^{a-\alpha}
                   \sum_{\epsilon=0}^c
                   {\cal P}{\cal A}(\theta,q)\,{\cal C}(\xpt, \ypt),
\end{equation}
where $c\equiv \alpha+\beta+\gamma+1$, 
\begin{equation}
{\cal P} = {(-1)^{\gamma} [ 1-(-1)^{\epsilon}]\over c} 2^{\alpha-\beta} 
           {a\choose\alpha} 
           {\alpha\choose\beta}
           {2b\choose\gamma} 
           {a-\alpha\choose\delta} 
           {c\choose\epsilon}, 
\end{equation}
\begin{displaymath}
{\cal A} = q^\epsilon (1-q^2)^{c-\epsilon}
           (\sin\theta)^{2(\alpha+\beta+b)-\epsilon+1} 
           (\cos\theta)^{2(\alpha+\gamma+\delta)-\epsilon+1}
\end{displaymath}
\begin{equation}
           ~~~~~~~ \qt^{-2c+\epsilon},
\end{equation}
and
\begin{equation}
{\cal C} =\ypt^{2(\alpha+b+\delta)-\epsilon+1}\,
          \xpt^{2(a-\alpha-\delta)} \,
          (1-\ltild^2 )^{\epsilon/2}, 
\end{equation}
with $\ell^2$ given in eq. (\ref{eq:ABqt}).

\subsection{Integration over isophotes}

In order to derive the quantity $\sigobsal$ we integrate
$\Sigtl\sigobs $ over isophotes, and this requires a bi--dimensional
integration of the quantities $I_{a,b}(\xp, \yp)$ in eq. (B.6).  From
eq. (B.7) the only factor affected by the integration is ${\cal C}$,
and so, with the natural parameterization:
\begin{equation}
\cases{
\xp = t\cos\lambda ,\cr
\yp = \qt t\sin\lambda ,
}
\end{equation}
according to eq. (\ref{eq:sigobsa}) we have 
\begin{equation}
\langle {\cal C}(\xpt ,\ypt) \rangle_\ell = \int_0^{2\pi}\!\!\!\!\int_0^\ell 
\!\!{\cal C} [{\tilde t}\cos\lambda,\qt {\tilde t}\sin\lambda ]
             \qt\,t\,dt\,d\lambda.
\end{equation}
The result of the integration can be rewritten as
\begin{equation}
\langle {\cal C}\rangle_\ell =\Rt^2\,
                              {\cal B}(\theta,q)\,
                              \Lambda_1(a,b,\alpha,\delta,\epsilon)\,
                              {\cal T}_1(a,b,\epsilon,\ltild),
\end{equation}
where 
\begin{equation}
{\cal B} =[1-(-1)^{\epsilon}]\,\qt^{2(\alpha+b+\delta+1)-\epsilon},  
\end{equation}
\begin{equation}
\Lambda_1 =\betaf\left(a-\alpha-\delta+{1\over 2},
           \alpha+b+\delta-{\epsilon\over 2}+1\right),  
\end{equation}
and
\begin{equation}
{\cal T}_1 ={\betaf_{a+b-\epsilon/2+ 3/2,\epsilon/2+1}(\ltild^2)\over 2}.
\end{equation}
Another quantity that appears in the expression of $\sigobsal$ is
$\langle \xp^2\,I_{a,b}(\xp ,\yp )\rangle_\ell$, due to  
the $\cos\varphi$ factor in eqs. (\ref{eq:vn}), and (\ref{eq:sign}). 
In this case, 
\begin{equation}
\langle\xp^2\,{\cal C}\rangle_\ell =\Rt^2 \,{\cal B}(\theta,q)\,
              \Lambda_2(a,b,\alpha,\delta,\epsilon)
              \,{\cal T}_2(a,b,\epsilon,\ltild), 
\end{equation}
where ${\cal B}$ is given in eq. (B.14), whit
\begin{equation}
\Lambda_2 =\betaf\left(a-\alpha-\delta+{3\over 2},
           \alpha+b+\delta-{\epsilon\over 2}+1\right),
\end{equation}
and
\begin{equation}
{\cal T}_2 ={\betaf_{a+b-\epsilon/2+5/2,\epsilon/2+1}(\ltild^2)\over 2}.
\end{equation}

\section{The $n=0$ model}

The properties of the constant density ellipsoid can obviously be
derived from the general expressions by setting $n=0$.  However, in
this special case it is less cumbersome to derive them directly from
their definition.  The total mass is given by
\begin{equation}
M={4\ \pi \,q\,\rhoc\,\Rt^3\over 3},
\end{equation}
while the surface density profile in the projection plane $(\xp ,\yp)$ is
given by
\begin{equation}
\Sigtl = {2\rhoc\Rt q\over \qt}\sqrt{1-\ltild^2}, 
\end{equation}
and the projected mass within isophote $\ell$ is
\begin{equation}
\Mtl ={\rhoc\Rt^3 4\pi q\over 3}[1-(1-\ltild^2)^{3/2}].
\end{equation}
From eq. (\ref{eq:elleff}), $\Reo\simeq 0.608\Rt$.
The potential inside the ellipsoid is
\begin{equation}
\phi(R,z) =-\pi G\rhoc\Rt^2 (\wz -\wu\Rtild^2 -\wt\ztild^2),
\end{equation}
where, from eq. (A5),
$\wz=\phi_{000}(q)=2q\arcsin\sqrt{1-q^2}/\sqrt{1-q^2}$, $\wu$ and 
$\wt$ are given in eqs. (A.11) and (A.12).
In case of nearly round galaxies, $\wz\sim 2(1+2q)/3 + {\rm
O}(1-q)^2$.  
Direct integration of eq. (\ref{eq:soljeans1}) shows that
\begin{equation}
\sigRs =\phi(R,\zt)-\phi(R,z)=\pi G\rhoc\Rt^2 q^2\wt (1-m^2):
\end{equation}
note how in this special case the radial velocity dispersion is
constant on the isodensity surfaces. 
From eq. (\ref{eq:soljeans2}) a simple algebra shows that
\begin{equation}
\tvphism -\sigRs=2\pi G\rhoc\Rt^2 (\wu -q^2\wt)\Rtild^2,
\end{equation}
where it can easily proved that $\wu -q^2\wt>0$ in the range $0 <q\leq
1$, and that for $q\to 1$ this quantity goes to zero as $\sim
8(1-q)/15$.  Equations (\ref{eq:Sigvp})--(\ref{eq:Sigsigp}) can be
easily evaluated and, from eqs. (C.5), (C.6), (\ref{eq:vn}) and
(\ref{eq:sign}), it follows that
\begin{equation}
\vp(\xp,\yp)=-\sqrt{2\pi G\rhoc\Rt^2 (\wu-q^2\wt)}\,k\,\xpt\sin\theta, 
\end{equation}
and
\begin{equation}
\Vps = \vp^2
\end{equation}
everywhere in the projection plane, and independently of $k$.
In addition, 
\begin{displaymath}
\sigps(\xp,\yp) = 2\pi G\rhoc\Rt^2
\end{displaymath}
\begin{equation}
~~~~ \left[{q^2\wt (1-\ltild^2)\over 3}+(1-k^2)(\wu
-q^2\wt)\xpt^2\sin^2\theta\right], 
\end{equation}
and, from eqs. (\ref{eq:Sigsiglos}) and (C.8)
\begin{equation}
\sigobs(\xp ,\yp)=\sigps(\xp, \yp).
\end{equation}
Note that from eq. (C.9) the quantity $\sigps (0,0)$ does not depend
on the inclination angle $\theta$, at variance with the cases $n >0$:
also the identities (C.8) and (C.10) are a peculiarity of the constant
density ellipsoid. 
Integration of $\sigobs$ over the isophote $\ell$ gives
\begin{equation}
\sigobsal ={\pi G\rhoc\Rt^2\over 5}\,\,
{{A_1+(1-k^2)\,\sin^2\theta\,A_2}\over{1-(1-\ltild^2)^{3/2}}}
\end{equation}
with
\begin{displaymath}
A_1= 2q^2\wt [1-(1-\ltild^2)^{5/2}],
\end{displaymath}
and
\begin{displaymath}
A_2 = (\wu -q^2\wt)[2-(2+\ltild^2-3\ltild^4)\sqrt{1-\ltild^2}].
\end{displaymath}

\end{document}